\documentclass[acmsmall,screen]{acmart}

\usepackage{listings}
\usepackage[utf8]{inputenc}
\usepackage[T1]{fontenc}
\usepackage{microtype}
\usepackage{soul}
\usepackage{multirow}
\usepackage{cleveref}
\usepackage{tabularx}

\crefname{section}{\S}{\S\S}
\crefname{figure}{Figure}{Figures}
\crefname{table}{Table}{Tables}

\lstdefinelanguage{obsidian}{
	keywords=[1]{contract,state,transaction,disown,new,asset,this,in,return,returns,remote,main,private,if,else,revert,implements,interface},
	keywordstyle=[1]\color{blue}\bfseries,
	keywords=[2]{int},
	keywordstyle=[2]\color{teal}\bfseries,
	keywords=[3]{->},
	keywordstyle=[3]\color{blue}\bfseries,
	keywords=[4]{Owned,Unowned,Shared},
	keywordstyle=[4]\color{black}\bfseries,
	comment=[l]{//},
	morecomment=[s]{/*}{*/},
	commentstyle=\color{gray}\ttfamily,
	stringstyle=\color{red}\ttfamily,
	morestring=[b]",
	alsoother={@},
}

\lstdefinelanguage{Solidity}{
	keywords=[1]{anonymous, assembly, assert, balance, break, call, callcode, case, catch, class, constant, continue, constructor, contract, debugger, default, delegatecall, delete, do, else, emit, event, experimental, export, external, false, finally, for, function, gas, if, implements, import, in, indexed, instanceof, interface, internal, is, length, library, log0, log1, log2, log3, log4, memory, modifier, new, payable, pragma, private, protected, public, pure, push, require, return, returns, revert, selfdestruct, send, solidity, storage, struct, suicide, super, switch, then, this, throw, transfer, true, try, typeof, using, value, view, while, with, addmod, ecrecover, keccak256, mulmod, ripemd160, sha256, sha3}, % generic keywords including crypto operations
	keywordstyle=[1]\color{blue}\bfseries,
	keywords=[2]{address, bool, byte, bytes, bytes1, bytes2, bytes3, bytes4, bytes5, bytes6, bytes7, bytes8, bytes9, bytes10, bytes11, bytes12, bytes13, bytes14, bytes15, bytes16, bytes17, bytes18, bytes19, bytes20, bytes21, bytes22, bytes23, bytes24, bytes25, bytes26, bytes27, bytes28, bytes29, bytes30, bytes31, bytes32, enum, int, int8, int16, int24, int32, int40, int48, int56, int64, int72, int80, int88, int96, int104, int112, int120, int128, int136, int144, int152, int160, int168, int176, int184, int192, int200, int208, int216, int224, int232, int240, int248, int256, mapping, string, uint, uint8, uint16, uint24, uint32, uint40, uint48, uint56, uint64, uint72, uint80, uint88, uint96, uint104, uint112, uint120, uint128, uint136, uint144, uint152, uint160, uint168, uint176, uint184, uint192, uint200, uint208, uint216, uint224, uint232, uint240, uint248, uint256, var, unit, ether, finney, szabo, wei, days, hours, minutes, seconds, weeks, years},	% types; money and time units
	keywordstyle=[2]\color{teal}\bfseries,
	keywords=[3]{block, blockhash, coinbase, difficulty, gaslimit, number, timestamp, msg, data, gas, sender, sig, value, now, tx, gasprice, origin},	% environment variables
	keywordstyle=[3]\color{violet}\bfseries,
	identifierstyle=\color{black},
	sensitive=false,
	comment=[l]{//},
	morecomment=[s]{/*}{*/},
	commentstyle=\color{gray}\ttfamily,
	stringstyle=\color{red}\ttfamily,
	morestring=[b]',
	morestring=[b]"
}

\newcommand{\code}[1]{\texttt{#1}}

\AtBeginDocument{%
  \providecommand\BibTeX{{%
    \normalfont B\kern-0.5em{\scshape i\kern-0.25em b}\kern-0.8em\TeX}}}

\setcopyright{rightsretained} 
\acmPrice{}
\acmDOI{10.1145/3428200}
\acmYear{2020}
\copyrightyear{2020}
\acmSubmissionID{oopsla20main-p23-p}
\acmJournal{PACMPL}
\acmVolume{4}
\acmNumber{OOPSLA}
\acmArticle{132}
\acmMonth{11}

\begin{document}


% Can advanced type systems be usable?
% Can ownership and typestate be usable? Then empirical study is a subtitle.
% Want concepts to be more general than just smart contracts.
\title[An Empirical Study of Obsidian]{Can Advanced Type Systems Be Usable? An Empirical Study of
Ownership, Assets, and Typestate in Obsidian}

\author{Michael Coblenz}
\orcid{0000-0002-9369-4069}             %% \orcid is optional
\affiliation{
  \position{Doctoral Candidate}
  \department{Computer Science Department}              %% \department is recommended
  \institution{Carnegie Mellon University}            %% \institution is required
  \streetaddress{5000 Forbes Ave.}
  \city{Pittsburgh}
  \state{PA}
  \postcode{15213}
  \country{USA}                    %% \country is recommended
}
\email{mcoblenz@cs.cmu.edu}          %% \email is recommended

%% Author with single affiliation.
\author{Jonathan Aldrich}
%\authornote{with author1 note}          %% \authornote is optional;
                                        %% can be repeated if necessary
\orcid{0000-0003-0631-5591}             %% \orcid is optional
\affiliation{
  \position{Professor}
  \department{Institute for Software Research}              %% \department is recommended
  \institution{Carnegie Mellon University}            %% \institution is required
  \streetaddress{5000 Forbes Ave.}
  \city{Pittsburgh}
  \state{PA}
  \postcode{15213}
  \country{USA}                    %% \country is recommended
}
\email{jonathan.aldrich@cs.cmu.edu}          %% \email is recommended

\author{Brad A. Myers}
%\authornote{with author1 note}          %% \authornote is optional;
                                        %% can be repeated if necessary
\orcid{0000-0002-4769-0219}             %% \orcid is optional
\affiliation{
  \position{Professor}
  \department{Human-Computer Interaction Institute}              %% \department is recommended
  \institution{Carnegie Mellon University}            %% \institution is required
  \streetaddress{5000 Forbes Ave.}
  \city{Pittsburgh}
  \state{PA}
  \postcode{15213}
  \country{USA}                    %% \country is recommended
}
\email{bam@cs.cmu.edu}          %% \email is recommended

%% Author with single affiliation.
\author{Joshua Sunshine}
%\authornote{with author1 note}          %% \authornote is optional;
                                        %% can be repeated if necessary
\orcid{0000-0002-9672-5297}             %% \orcid is optional
\affiliation{
  \position{Systems Scientist}
  \department{Institute for Software Research}              %% \department is recommended
  \institution{Carnegie Mellon University}            %% \institution is required
  \streetaddress{5000 Forbes Ave.}
  \city{Pittsburgh}
  \state{PA}
  \postcode{15213}
  \country{USA}                    %% \country is recommended
}
\email{joshua.sunshine@cs.cmu.edu}          %% \email is recommended

\renewcommand{\shortauthors}{Coblenz et al.}

\begin{abstract}
Some blockchain programs (smart contracts) have included serious security vulnerabilities. Obsidian is a new typestate-oriented programming language that uses a strong type system to rule out some of these vulnerabilities. Although Obsidian was designed to promote \textit{usability} to make it as easy as possible to write programs, strong type systems can cause a language to be difficult to use. In particular, ownership, typestate, and assets, which Obsidian uses to provide safety guarantees, have not seen broad adoption together in popular languages and result in significant usability challenges. We performed an empirical study with 20 participants comparing Obsidian to Solidity, which is the language most commonly used for writing smart contracts today. We observed that Obsidian participants were able to successfully complete more of the programming tasks than the Solidity participants. We also found that the Solidity participants commonly inserted asset-related bugs, which Obsidian detects at compile time.
\end{abstract}

%% 2012 ACM Computing Classification System (CSS) concepts
%% Generate at 'http://dl.acm.org/ccs/ccs.cfm'.
\begin{CCSXML}
<ccs2012>
<concept>
<concept_id>10011007.10011006.10011008.10011024</concept_id>
<concept_desc>Software and its engineering~Language features</concept_desc>
<concept_significance>500</concept_significance>
</concept>
<concept>
<concept_id>10011007.10011006.10011050.10011017</concept_id>
<concept_desc>Software and its engineering~Domain specific languages</concept_desc>
<concept_significance>500</concept_significance>
</concept>
<concept>
<concept_id>10002978.10003022</concept_id>
<concept_desc>Security and privacy~Software and application security</concept_desc>
<concept_significance>300</concept_significance>
</concept>
<concept>
<concept_id>10003120.10003121.10011748</concept_id>
<concept_desc>Human-centered computing~Empirical studies in HCI</concept_desc>
<concept_significance>300</concept_significance>
</concept>
</ccs2012>
\end{CCSXML}

\ccsdesc[500]{Software and its engineering~Language features}
\ccsdesc[500]{Software and its engineering~Domain specific languages}
\ccsdesc[300]{Security and privacy~Software and application security}
\ccsdesc[300]{Human-centered computing~Empirical studies in HCI}
%% End of generated code

\keywords{typestate, linear types, ownership, assets, permissions, blockchain, smart contracts, empirical studies of programming languages}

\maketitle

\section{Introduction}
\label{sec:introduction}

\begin{sloppypar}
Obsidian~\cite{Coblenz2020:Obsidian} is a new programming language for writing \textit{smart contracts}~\cite{Szabo97:Formalizing}, which are programs for blockchain platforms. Blockchains aim to provide a trusted computing medium among users who have not necessarily established trust. By decentralizing computation, system designers can obtain strong security properties. Unfortunately, these properties can only be obtained when the smart contracts themselves implement the intended behavior. Through bugs and security vulnerabilities in blockchains, attackers have stolen millions of dollars worth of cryptocurrencies~\cite{DAO, CNBC}.
\end{sloppypar}

To improve safety and prevent vulnerabilities, researchers have developed new languages with stronger safety properties. In this work, we focus on Obsidian, which stands for \textit{Overhauling Blockchains with States to Improve Development of Interactive Application Notation}. Obsidian was based on a requirements analysis for smart contract tools~\citep{Coblenz2019:Smarter} to provide safety properties that would be relevant to many blockchain applications. In addition, Obsidian was designed to be \textit{usable} through iterative user testing with 44 participants~\cite{Coblenz2020:PLIERS} so that programmers would be able to use it effectively. In prior work, we showed that most of the participants in a six-participant qualitative study were able to complete relevant programming tasks~\cite{Coblenz2020:PLIERS}. However, that small study did not compare Obsidian to any other languages. 

Ours is the first quantitative user study (of which we are aware) of a type system that supports linear types, ownership, or typestate, or any combination of these together. Earlier formative studies of Obsidian isolated particular language features, sometimes by adapting those features in the context of a language with which participants were already familiar. In contrast, in this summative study we studied the full Obsidian language, including the compiler and editor. Our study compares Obsidian to the most widely-used language for writing smart contracts,  \textit{Solidity}~\cite{Solidity}.

Although Solidity and Obsidian target different blockchain platforms (Ethereum \cite{Ethereum} and Hyperledger Fabric \cite{Fabric}, respectively), we chose Solidity for comparison because it is the dominant programming language used in public smart contract development and because it is the language that was used to develop popular smart contracts that had serious security vulnerabilities~\cite{DAO, CNBC}.

\begin{figure}[t]
\begin{lstlisting}[language=Obsidian,xleftmargin=.2\linewidth]
asset contract Medicine {}
asset contract Pharmacy 
{
  Medicine@Owned med;
  transaction getNewMedicine(Medicine@Owned >> Unowned m) 
  { 
    med = m;
  }
}
\end{lstlisting}
\centering
\includegraphics[width=.7\columnwidth]{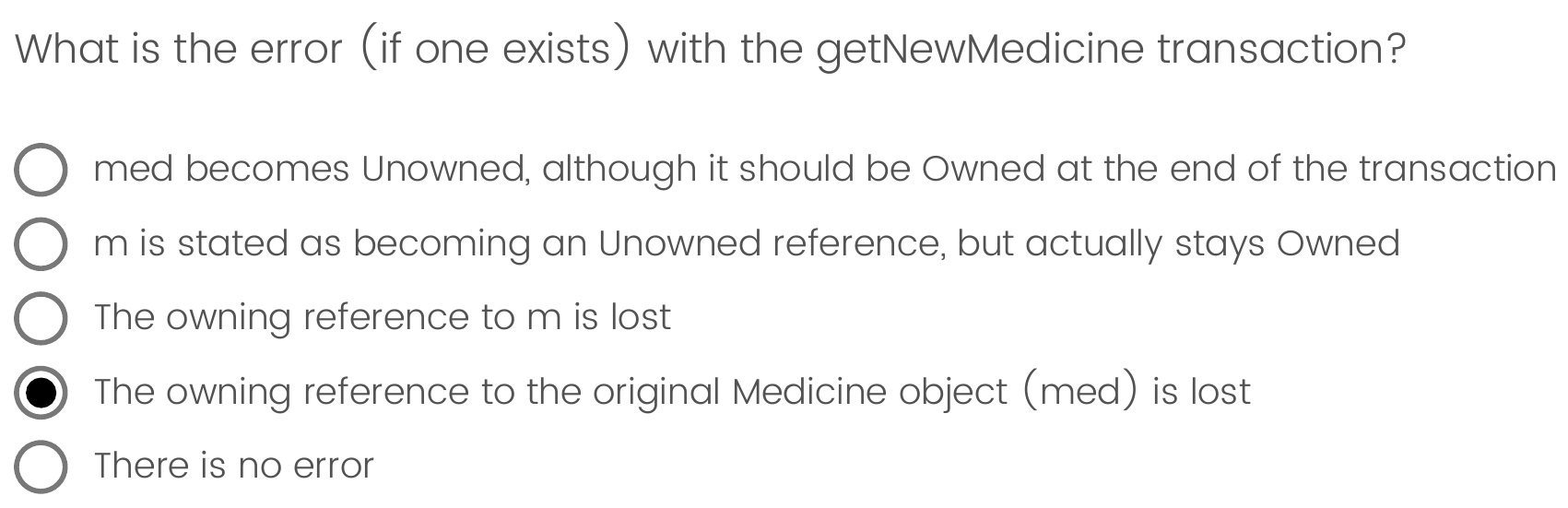}
\caption{An example practice question on assets from the Obsidian tutorial (showing the correct answer).}
\label{tutorial-example}
\Description{The practice question gives the following code: 
asset contract Medicine {}
asset contract Pharmacy 
{
  Medicine@Owned med;
  transaction getNewMedicine(Medicine@Owned >> Unowned m) 
  { 
    med = m;
  }
}

Then, it asks, "What is the error (if one exists) with the getNewMedicine transaction?" The choices are "med becomes Unowned, although it should be Owned at the end of the transaction"; "m is stated as becoming an Unowned reference, but actually stays Owned"; "The owning reference to m is lost"; "The owning reference to the original Medicine object (med) is lost"; "There is no error". The second to last option ("The owning reference to the original Medicine object (med) is lost") is marked as being correct.}
\end{figure}

Considering Obsidian's design goal to improve safety with a novel type system, we identified three high-level research questions for our study:

\begin{description}
    \item[RQ1:] Could we obtain actionable data about the usability of a novel programming language (that uses a type system that would be unfamiliar to our participants) in a short-duration user study (less than one day), which would be representative of real-world smart contract development?
    \item[RQ2:] Do programmers using Solidity insert more of the kinds of bugs that Obsidian is designed to catch?
    \item[RQ3:] Strong type systems can impose a usability burden on programmers because the compiler forces programmers to write code so that the compiler can verify various safety properties. Can programmers who were previously familiar with object-oriented programming (but not with Obsidian, typestate, ownership, or linear type systems in general) successfully use Obsidian to complete relevant smart contract programming tasks? If so, is there a significant impact on task completion times?
\end{description}

In RQ1, \textit{actionable data} refers to data that inform programming language designers about their designs in ways that have implications on language design or language training. We used these three research questions to develop task criteria (\cref{sec:criteria}). Then, we developed three tasks according to the criteria: Auction (\cref{sec:auction}), Prescription (\cref{sec:prescription}), and Casino (\cref{sec:casino}). In each task, we refined these high-level research questions into task-relevant ones.

In summary, after the training period (about 90 minutes), seven of the ten Obsidian participants were able to successfully use Obsidian to finish implementing the required small program in our Auction task. In contrast, only two of the ten Solidity participants finished the task correctly. In the Prescription task, seven of the ten Obsidian participants were able to fix a security vulnerability in the program, whereas only two of the ten Solidity participants were able to do so. Six of the Obsidian participants used ownership to do the task, suggesting that ownership is teachable in a relatively short training period.

Five of the Obsidian participants had enough time remaining to do the Casino task (meaning that the \textit{other} participants felt that their solutions were incomplete when their four-hour time expired). Although four of these wrote solutions that compiled, all four abused the \code{disown} operator, resulting in asset loss. Among the eight Solidity participants who had enough time for this task, half inserted bugs that resulted in improper asset fabrication or loss, showing that it is still important to pursue approaches to prevent asset abuse. Only one participant, who was in the Solidity condition, finished the task correctly.

After the tasks had ended, we gave participants a post-study survey asking for their opinions about the language they had used. We found that Obsidian participants felt that ownership was useful. They also observed that the tutorial and exercises were effective tools for learning the language.

We come to three conclusions in this paper. First, the methodology we used, in which we include practice-based training in a traditional quantitative study, is effective for evaluating a novel programming language. Second, despite the strong, unfamiliar restrictions it imposes, Obsidian's ownership and permissions system can be taught to some kinds of programmers in a short period of time in a way that results in many of them being able to use it effectively. That is, for some tasks, Obsidian enables people to be more successful at writing code that is free of the serious bugs that Obsidian detects. Third, without using techniques to detect or prevent asset-loss bugs, when doing some kinds of smart contract programming tasks, programmers will accidentally insert these kinds of bugs.

\section{The Obsidian Language}
Obsidian uses \textit{ownership} to express in the type system that each \textit{asset} (an object that has value and therefore should not be lost) has exactly one owning reference. Ownership represents a form of \textit{linear types}~\cite{Wadler90:Linear}. Ownership may be transferred between references, but if the owning reference goes out of scope, then the compiler reports an error. This ensures that owning references to assets cannot be lost unless they are explicitly \code{disown}ed by the programmer. This design was motivated by \citet{Delmolino2016:Step}, who found in a user study that programmers of smart contracts inserted bugs that lost assets.

Obsidian extends ownership with \textit{typestate}~\cite{strom1986typestate} because blockchain programs are typically stateful with objects supporting different operations depending on their state \cite{Solidity-Patterns-States}. Thus, owning references can also specify what \textit{state} the referenced object is in. For example, \code{Auction@Open} is the type of a reference to an \code{Auction} object that is in state \code{Open}. Because only owning references can specify typestate in Obsidian, ownership is implied by the presence of a typestate specification. Alternatively, \code{Auction@Owned} is the type of an owned reference to an \code{Auction} object, which may be in any state. Ownership is one example of \textit{linear} types~\cite{Wadler90:Linear}: types that are consumed when used rather than being duplicated, and which \textit{must} be consumed rather than merely dropped.

To improve flexibility, Obsidian also allows \code{Shared} references, which can refer to an object that has no owner, and \code{Unowned} references, which refer to an object that may have an owner. These additional options allow users to avoid establishing ownership structures when none are needed. Unlike \code{Owned} and \code{Shared} references, \code{Unowned} references cannot be used to change which of the various named states the object is in. This is required for soundness, since a change in state through an \code{Unowned} reference might violate a typestate specification on the owning reference. These annotations (\code{Owned}, \code{Shared}, and \code{Unowned}) are referred to as \textit{permissions} because they denote what operations can be done via references with those annotations.

Obsidian is a class-based object-oriented language that uses a syntax similar to Java. Obsidian uses the keyword \code{contract} instead of \code{class} due to conventions of blockchain platforms, and calls methods \code{transaction}s because invocations of blockchain code from outside the blockchain have transactional semantics: either the execution finishes successfully and new state is stored in the blockchain, or the transaction is reverted and any state changes are not preserved. Currently, Obsidian supports the Hyperledger Fabric blockchain platform~\cite{Hyperledger}.

We provided a full description and formal treatment of Obsidian~\citep{Coblenz2020:Obsidian}; here, we explain Obsidian by example. The left column of \cref{obsidian-auction-task-code} shows part of an \code{Auction} contract. An \code{Auction} instance is always in one of three states: \code{Open}, \code{BidsMade}, or \code{Closed}. A field, \code{seller}, is always in scope, but when the object is in state \code{BidsMade}, fields \code{maxBidder} and \code{maxBid} are also in scope. Transactions specify types for their parameters, but in addition to the normal parameters, when \code{this} is specified as the first parameter, the program can specify a type for the \code{this} reference. The \code{bid} transaction, for example, can only be invoked via a \code{Shared} reference to the receiver. Formal parameters use \code{>>} to denote changes in permission or state. For example, the \code{money} parameter of \code{bid} is specified with permission \code{Owned >> Unowned}, meaning that the caller must pass an \code{Owned} reference, and when \code{bid} returns, from the perspective of the caller, the \code{money} is now \code{Unowned}. In other words, ownership was passed from the caller to the transaction.

\begin{figure*}[ht]
\begin{minipage}[t]{0.48\linewidth}
\lstset{numbers=left, belowskip=0pt, basicstyle=\ttfamily\tiny, xleftmargin=11pt, numbersep=5pt}
\begin{lstlisting}
main asset contract Auction {
  Participant@Unowned seller;

  state Open;
  state BidsMade {
    // the bidder who made the highest bid so far
    Participant@Unowned maxBidder; 
    Money@Owned maxBid;
  }
  state Closed;

  (*@\ldots@*)




  transaction bid(Auction@Shared this, 
                  Money@Owned >> Unowned money, 
                  Participant@Unowned bidder) {
      if (this in Open) {
        // Initialize destination state, 
        // and then transition to it.
        BidsMade::maxBidder = bidder;
        BidsMade::maxBid = money;
        ->BidsMade;
      }
      else {
        if (this in BidsMade) {
          //if the new bid  > current Bid
          if (money.getAmt() > maxBid.getAmt()) { 
            //1. TODO: fill this in. 
            //Can call other transactions.
  \end{lstlisting}
  \begin{lstlisting}[backgroundcolor=\color{yellow}, numbers=left, firstnumber=last, aboveskip=0pt, belowskip=0pt, basicstyle=\ttfamily\tiny]
            maxBidder.receivePayment(maxBid);
            maxBidder = bidder;
            maxBid = money;
  \end{lstlisting}
  \begin{lstlisting}[numbers=left, firstnumber=last, aboveskip=0pt, belowskip=0pt, basicstyle=\ttfamily\tiny]
          }
          else {
            //2. TODO: return money to the bidder, 
            //   since the new bid was too low.
            //Can call other transactions.
  \end{lstlisting}
  \begin{lstlisting}[backgroundcolor=\color{yellow}, numbers=left, firstnumber=last, aboveskip=0pt, belowskip=0pt, basicstyle=\ttfamily\tiny]
            bidder.receivePayment(money);
  \end{lstlisting}
  \begin{lstlisting}[numbers=left, firstnumber=last, aboveskip=0pt, belowskip=0pt, basicstyle=\ttfamily\tiny]
          }
        }
        else {
          revert("Can't bid on closed auctions.");
        }
      }
    }
}
\end{lstlisting}
\caption*{Obsidian condition.}
\end{minipage}
\hspace{0.3cm}
\begin{minipage}[t]{0.48\linewidth}
\lstset{numbers=none, belowskip=0pt, basicstyle=\ttfamily\tiny}
\begin{lstlisting}[language=Solidity, belowskip=0pt, basicstyle=\ttfamily\tiny]
contract Auction {
  // the bidder who made the highest bid so far
  address maxBidder; 
  uint maxBidAmount;
  
  // 'payable' indicates we can transfer money 
  // to this address
  address payable seller; 

  // Allow withdrawing previous money 
  // for bids that were outbid
  mapping(address => uint) pendingReturns;
  
  enum State { Open, BidsMade, Closed }
  State state;
  (*@\ldots@*)
  function bid() public payable {


    if (state == State.Open) {
      maxBidder = msg.sender;
      maxBidAmount = msg.value;
      state = State.BidsMade;
    }
    
    
    else {
      if (state == State.BidsMade) {
  	    //if the new bid  > current Bid
        if (msg.value > maxBidAmount) {
          //1. TODO: fill this in. 
          //Can call other functions as needed.
  \end{lstlisting}
  \begin{lstlisting}[backgroundcolor=\color{yellow}, aboveskip=0pt, belowskip=0pt, basicstyle=\ttfamily\tiny]
          pendingReturns[maxBidder] += maxBidAmount;
          maxBidder = msg.sender;
          maxBidAmount = msg.value;
  \end{lstlisting}
  \begin{lstlisting}[aboveskip=0pt, belowskip=0pt, basicstyle=\ttfamily\tiny]
        }
        else {
          //2. TODO: return money to the bidder, 
          //   since the new bid was too low.
          //Can call other functions as needed.
  \end{lstlisting}
  \begin{lstlisting}[backgroundcolor=\color{yellow}, aboveskip=0pt, belowskip=0pt, basicstyle=\ttfamily\tiny]
          pendingReturns[msg.sender] += msg.value;
  \end{lstlisting}
  \begin{lstlisting}[aboveskip=0pt, belowskip=0pt, basicstyle=\ttfamily\tiny]
        }
      }
      else {
        revert("Can't bid on closed auctions.");
      }
    }
  }
}
\end{lstlisting}
\caption*{Solidity condition.}
\end{minipage}
\caption{A side-by-side example comparing Obsidian code to Solidity code, taken from the two conditions of the Auction task (with minor changes to fit on this page). Code highlighted in yellow represents a correct solution; the rest was given to participants as starter code. In the Obsidian code, line 33 transfers ownership of the object referenced by \code{maxBid} to the \code{receivePayment} parameter. The new type of \code{maxBid} from then until line 35 is \code{Money@Unowned}. Line 35 re-establishes ownership in \code{maxBid} by transferring ownership from \code{money} to \code{maxBid}.}
\label{obsidian-auction-task-code} 
\Description{
Obsidian condition:

    main asset contract Auction {
      Participant@Unowned seller;
    
      state Open;
      state BidsMade {
        // the bidder who made the highest bid so far
        Participant@Unowned maxBidder; 
        Money@Owned maxBid;
      }
      state Closed;
    
      ...
    
      transaction bid(Auction@Shared this, 
                      Money@Owned >> Unowned money, 
                      Participant@Unowned bidder) {
          if (this in Open) {
            // Initialize destination state, 
            // and then transition to it.
            BidsMade::maxBidder = bidder;
            BidsMade::maxBid = money;
            ->BidsMade;
          }
          else {
            if (this in BidsMade) {
              //if the new bid  > current Bid
              if (money.getAmt() > maxBid.getAmt()) { 
                //1. TODO: fill this in. 
                //Can call other transactions as needed.
                // SAMPLE SOLUTION:
                maxBidder.receivePayment(maxBid);
                maxBidder = bidder;
                maxBid = money;
              }
              else {
                //2. TODO: return money to the bidder, 
                //   since the new bid was too low.
                //Can call other transactions as needed.
                // SAMPLE SOLUTION:
                bidder.receivePayment(money);
              }
            }
            else {
              revert("Can't bid on closed auctions.");
            }
          }
        }
    }
Solidity condition:
    contract Auction {
      // the bidder who made the highest bid so far
      address maxBidder; 
      uint maxBidAmount;
      
      // 'payable' indicates we can transfer money 
      // to this address
      address payable seller; 
    
      // Allow withdrawing previous money 
      // for bids that were outbid
      mapping(address => uint) pendingReturns;
      
      enum State { Open, BidsMade, Closed }
      State state;
      ...
      function bid() public payable {
    
        if (state == State.Open) {
          maxBidder = msg.sender;
          maxBidAmount = msg.value;
          state = State.BidsMade;
        }

        else {
          if (state == State.BidsMade) {
      	    //if the new bid  > current Bid
            if (msg.value > maxBidAmount) {
              //1. TODO: fill this in. 
              //Can call other functions as needed.
			  // SAMPLE SOLUTION:
              pendingReturns[maxBidder] += maxBidAmount;
              maxBidder = msg.sender;
              maxBidAmount = msg.value;
            }
            else {
              //2. TODO: return money to the bidder, 
              //   since the new bid was too low.
              //Can call other functions as needed.
			  // SAMPLE SOLUTION:
              pendingReturns[msg.sender] += msg.value;
            }
          }
          else {
            revert("Can't bid on closed auctions.");
          }
        }
      }
    }
}
\end{figure*}

Dynamic state tests, via the \code{in} operator, check the dynamic state of a referenced object. For example, line 20 checks to see if \code{this} is in state \code{Open}. If the test passes, lines 23-24 initialize the \code{maxBidder} and \code{maxBid} fields of the \code{BidsMade} state, to which the object transitions on line 25. Line 24 transfers ownership of an object initially referenced by \code{money} to the \code{maxBid} field, leaving \code{money} with type \code{Money@Unowned}. 

Fields can temporarily have types that differ from their declarations as long as by the end of the transaction, all fields have types consistent with their declarations. For example, line 31 returns the money from the previous \code{maxBidder} which causes field \code{maxBid} to temporarily have type \code{Money@Unowned}; this is corrected on line 33, which transfers ownership from \code{money} to \code{maxBid}.

The \code{revert} statement (line 45) discards all changes that have been made to state in the transaction and reports an error.

\section{The Solidity Language}
Solidity~\cite{Solidity}, like Obsidian, is a class-based object-oriented language. Solidity targets the Ethereum blockchain platform~\cite{Ethereum}. The \code{function} keyword denotes methods, though methods have transactional semantics. It has no built-in notion of states, but programs can declare \code{enum}s and use them to represent states. Solidity supports a built-in cryptocurrency called \textit{ether}. Functions that are annotated \code{payable} can receive quantities of ether; the ether is conceptually sent with the invocation, and the amount is stored in the variable \code{msg.value}. Each contract instance can own a quantity of ether; this quantity is automatically updated by the runtime when a function receives a payment. Every contract instance is stored on the blockchain at a particular address. The language has a built-in type called \code{address} to represent these addresses.

Restricting only functions annotated \code{payable} to receive ether may prevent some kinds of lost-ether bugs, in which clients accidentally send ether to the wrong function. However, this approach does not apply to other kinds of resources, and only covers \textit{whether} the function can receive ether, not whether the function body accounts for it correctly. Likewise, although Solidity includes support for permission modifiers, such as \code{private}, to protect sensitive functions from being called externally, these modifiers do not prevent the bodies of methods from abusing assets or from being invoked when the receiving contract is in an inappropriate state.

Programs typically implement their own fine-grained accounting mechanism. For example, the \code{pendingReturns} structure records how much money is owed to each of a number of addresses. Without this mapping, although the contract would still record how much ether it held, the implementation would not be able to track for whom it is being kept.

The \code{pendingReturns} mapping supports the \textit{withdrawal pattern}~\cite{Withdrawal-Pattern}, which is an Ethereum coding convention that protects against re-entrancy attacks. The possibility of attacks arises because sending ether to a contract can cause the recipient to execute arbitrary code: the recipient can invoke a function on the sender, to which there is already an invocation on the stack. This is dangerous if the funds were sent while the sender was in an inconsistent state. Instead, it is recommended that contracts merely record that they owe ether to the intended recipient and provide a \code{withdraw} function that recipients can call to retrieve their money. The withdrawal pattern is not used in Obsidian, since Obsidian targets Hyperledger Fabric, which does not have this vulnerability. Because this was a summative study of the programming environments that users would encounter when using each language, we expected participants to use the withdrawal pattern with Solidity and not with Obsidian. This introduced the risk that the use of the withdrawal pattern would cause additional complexity relative to Obsidian.

In Solidity, using the withdrawal pattern typically results in the programmer writing code with arithmetic operations to update balances. In contrast, an Obsidian implementation of the withdrawal pattern would be expected to use linear assets, which the compiler could check for abuse --- making the withdrawal pattern safer on Obsidian than Solidity. Although relevant to this study, if Obsidian were used with Ethereum, we expect Obsidian's asset-based approach would guard against bugs in using the withdrawal pattern, regardless whether the opportunity for it arises from the platform design or from the particular API being used.

\section{Study Design}

Our experimental protocol was approved by our university IRB. The experiment began with obtaining informed consent. Participants (\cref{sec:participants}), whom we recruited from the university community, were randomly assigned to use either Obsidian or Solidity (using a block size of two) and given a tutorial (\cref{sec:training}) on their assigned language, during which an experimenter answered any questions they had. Then, the study proceeded with three tasks: Auction (\cref{sec:auction}), Prescription (\cref{sec:prescription}), and Casino (\cref{sec:casino}). After the tasks had ended, we gave participants a post-study survey (\cref{sec:post-study-survey}) asking for their opinions about the language they had used. We compensated participants with a \$75 Amazon gift card.

Our participants had a variety of levels of programming skill and were new at programming in Obsidian and mostly new at programming in Solidity. To make it more likely that they would complete at least some tasks and to leverage the skills that participants would acquire by doing the tasks, we gave the tasks to the participants in order of increasing difficulty (according to our experience with pilot studies). All participants worked on the tasks in the same order.

Prior empirical work in programming language evaluation found that testing and debugging programs in the context of empirical studies substantially increases variance. For example, we previously found that different participants have different levels of thoroughness in writing tests and different levels of debugging skill~\citep{Coblenz2017:Glacier}. When allowed to test and debug, participants frequently spend large amounts of time debugging issues that are not relevant to the experiment. To focus our participants' time on work related to our research questions, we allowed them to edit their code until they were satisfied, but did not give them an opportunity to test their code. Then, instead of assessing programs on the basis of tests, we inspected their code. We looked both for specific bugs that corresponded to the research questions we were interested in as well as for unrelated bugs. To do this, we developed a rubric for each task, which listed particular bugs to look for; we iterated on this rubric to add bugs that were present in the programs. This approach was made feasible by the relatively small codebases (see \cref{program-lengths}). Although it is possible that we missed particular bugs in the code, the same would be true even if we provided unit tests. However, by evaluating by inspection rather than unit tests, we were able to assess code with significant functional problems, similar to how one might grade a student exam by using a rubric rather than by executing unit tests. By using a rubric, partial credit can be awarded and specific mistakes can be identified even when the solution contains significant flaws.

Our decision to allow compilation but not testing was also motivated by our desire to evaluate the ability of Obsidian's type system to detect bugs that might otherwise be introduced. In many cases, it is cheaper to find bugs earlier in the software development process than later, so we focused on whether the kinds of bugs that Obsidian can detect at compile time are ones that are introduced at all and how successfully Obsidian programmers can complete tasks. In \citet{Delmolino2016:Step}, the programmers had access to a blockchain and still finished their work with code that lost assets; nonetheless, this does reflect a limitation in the study design. Because Obsidian's design is focused on identifying more bugs at \textit{compile time}, allowing compilation but not testing allowed us to focus on our research questions about the usability and effectiveness of the \textit{type system} while using our participants' time as effectively as possible. However, this may have introduced a kind of bias, since the Solidity participants might have found some of their bugs through testing if that had been permitted.

Another approach that could improve external validity is adding code review to the process. We did not include an explicit code review step in which peers inspected the code; although this might more-realistically represent some settings, studies have found that code reviews only find bugs infrequently~\cite{Czerwonka2015:Code}. Also, adding code reviews to the process would have substantially increased time requirements as well as variance, since the quality of the feedback would necessarily have varied.

\section{Participants}
\label{sec:participants}

We recruited participants to our four-hour study with posters at our university, with emails to lists of appropriate degree programs (such as the Master of Software Engineering program), and by advertising to students in relevant courses (e.g., a software engineering course). 

Because we advertised the study broadly (the posters were visible to anyone on campus), and because the study assumed that participants already knew an object-oriented programming language, we wanted to make sure that all participants had appropriate preparation. Therefore, we pre-screened participants with an online survey that asked them basic Java questions; we invited respondents who answered five of six questions correctly to participate in the study. The complete pre-screening instrument is in the supplement as well as in the replication package~\cite{Coblenz2020:ObsidianReplication}. The six questions concerned: Java constructor syntax; the definition of encapsulation; whether changes to a list through a reference would be visible through another reference; whether methods in interfaces may include bodies; whether abstract classes may be instantiated; and whether concrete subclasses of abstract classes must implement methods that were abstract in the superclass.

The same pre-screening instrument was used for earlier Obsidian studies in addition to this experiment. Over the period during which the instrument was used, 53 people completed it. Scores on the ``basic Java'' portion were out of six. Of the 53 respondents, seven scored below five, 13 scored five, and 33 scored six. Of those who participated in the RCT described in this paper, six scored five and the remaining participants scored six. Of the respondents who were not recruited in earlier studies, we contacted those with scores of at least five in the order in which they filled out the instrument. Those who responded to our requests were scheduled according to their and the experimenters' schedule requirements. No participants left the study early.

\Cref{participant-experience} summarizes the previous experience of the experiment participants in each condition. We excluded an additional participant who took so long on the training phase that not enough time was available for more than one programming task\footnote{That participant spent 3 hours and 11 minutes on the tutorial, which was three standard deviations above the mean.}. This left 20 participants; 14 of them identified as male, and six as female. Only two participants, both in the Obsidian condition, had no professional experience. Blockchains are targeted at enabling general software engineers in a wide variety of industries to build applications in contexts that lack mutual trust. Since our participants had substantial programming experience and 18 of them had some professional experience, we argue that our participants were representative of at least some kinds of programmers in industry who might be interested in writing smart contracts.

% Excluded P54.

\begin{table}[b]
\begin{tabular}{@{}lll@{}}
\toprule
& \textbf{Solidity ($N = 10$)} & \textbf{Obsidian ($N = 10$)} \\
\midrule
Median programming experience, years (range) & 9.2 (3.3 - 13) & 5.0 (2.3 - 8.5) \\
Median professional experience, years (range) & 1 (0.5-9.0) & 0.92 (0.0 - 5.0) \\
Median Java experience, years (range) & 2.0 (0.67 - 4.2) & 1.5 (0.25 - 6.0) \\
\bottomrule
\end{tabular}
\caption{Participant experience (self-reported).}
\label{participant-experience}
\end{table}

\section{Training}
\label{sec:training}

We provided a web-based tutorial (implemented with Qualtrics, and included in the supplement), which stepped participants through web-based documentation and exercises for their assigned language. Some of the exercises were to be completed in Visual Studio Code, which was configured with a compiler. Some of the questions were multiple-choice; for these, the tool automatically showed participants if they had entered an incorrect answer. An experimenter was available to answer questions. Participants were told that they should try to get all of their questions addressed during the training phase, since no questions could be answered after training was completed.

\begin{figure}[tb]
\fbox{\includegraphics[width=.97 \columnwidth]{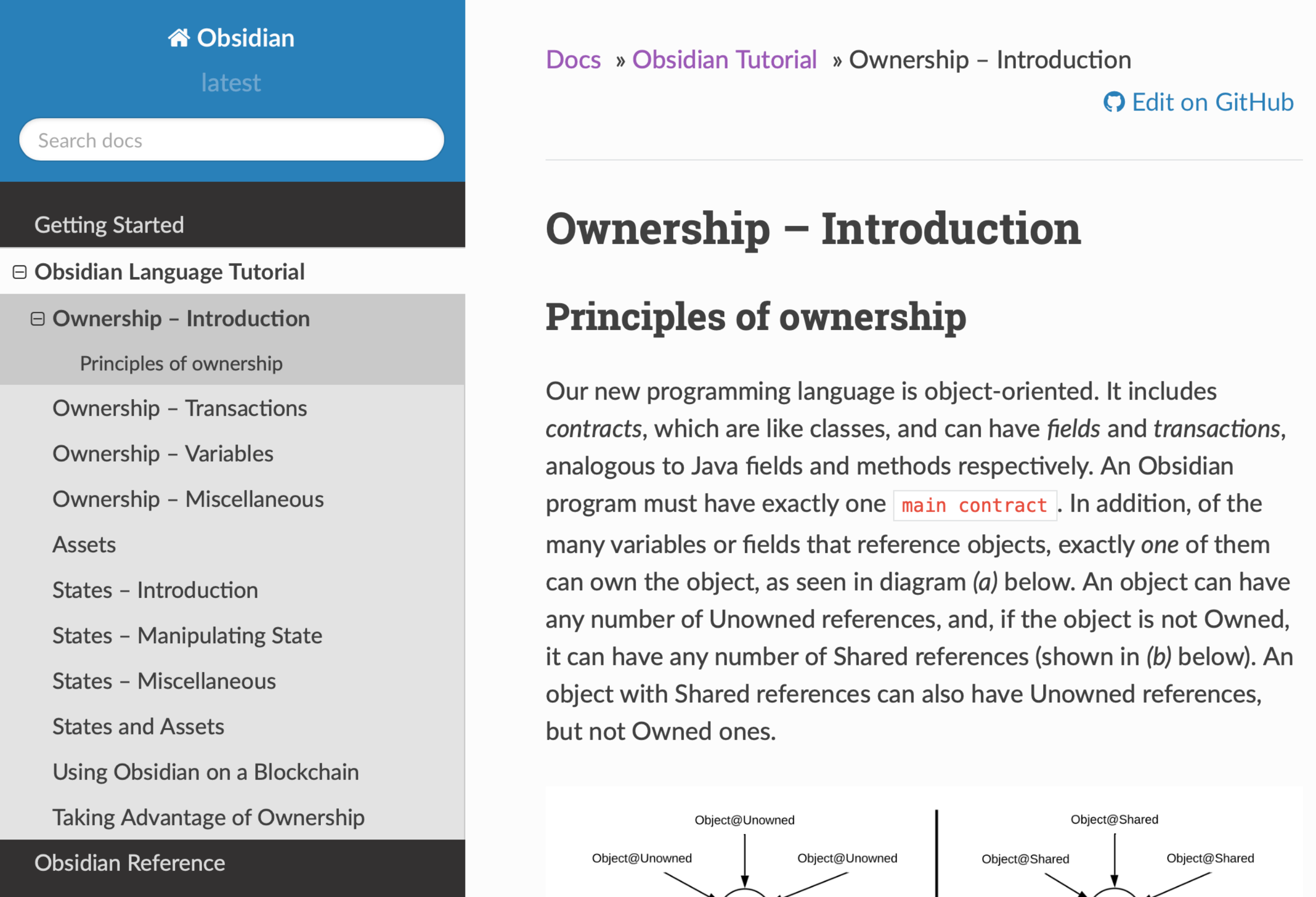}}
\caption{An example from the web-based Obsidian tutorial.}
\label{obsidian-docs}
\Description{The web page shows a column on the left with sections of the tutorial. The main section shows the "Ownership -- Introduction" section, with heading "Principles of ownership." Other categories include "Ownership -- Transactions," "Ownership -- Variables," "Ownership -- Miscellaneous," "Assets," "States -- Introduction", "States -- Manipulating State," "States -- Miscellaneous," "States and Assets," "Using Obsidian on a Blockchain," and "Taking Advantage of Ownership." The text in the main section is "Our new programming language is object-oriented. It includes contracts, which are like classes, and can have fields and transactions, analogous to Java fields and methods respectively. An Obsidian program must have exactly one main contract. In addition, of the many variables or fields that reference objects, exactly one of them can own the object, as seen in diagram (a) below. An object can have any number of Unowned references, and, if the object is not Owned, it can have any number of Shared references (shown in (b) below). An object with Shared references can also have Unowned references, but not Owned ones."}
\end{figure}

We included practice questions in the tutorial to ensure that participants absorbed the material (prior studies had found that without practice questions, participants skimmed the material without mastering the concepts). \Cref{tutorial-example} shows an example practice question. \Cref{obsidian-docs} shows a sample from the Obsidian documentation.

To make the two experimental conditions as similar as possible, even though Solidity includes no support for ownership, typestate, or assets, Solidity participants received a tutorial that explained these concepts and recommended using comment-based annotations. If participants asked about the utility of these annotations, we argued that this was similar to how one might write preconditions or postconditions in comments. \Cref{training-times} summarizes the distribution of times participants spent on the tutorial in the two conditions. If we had trained only the Obsidian participants in ownership, for example, any differences in behavior could have been attributed to the training and not to the language they were using.

Solidity participants received training on the \textit{withdrawal pattern} and its necessity for security reasons, as well as language-specific details such as the \code{payable} keyword and the concept of \textit{ether}.

\begin{table}[tb]
\begin{tabular}{@{}lll@{}}
\toprule
& \textbf{Solidity ($N = 10$)} & \textbf{Obsidian ($N = 10$)} \\
\midrule
Average (standard deviation) & 86 (28) min. & 98 (31) min. \\
Range & 39 to 138 min. & 50 to 148 min. \\
\bottomrule
\end{tabular}
\caption{Training times in Solidity and Obsidian conditions.}
\label{training-times}
\end{table}
%T test results:
%t = 0.96039, df = 17.879, p-value = 0.3497
%alternative hypothesis: true difference in means is not equal to 0
%95 percent confidence interval:
% -15.09553  40.49553

\section{Task Selection}
\label{sec:criteria}

Based on our research questions, we sought a collection of tasks for our study that met several criteria. We selected criteria, shown in \cref{table:task-design-criteria}, to help us identify tasks that would address our research questions. 

\begin{table}[ht]
    \centering
    \begin{tabularx}{\linewidth}{@{}l X lp{2cm}@{}}
        \toprule
        \textbf{Name} & \textbf{Criterion} & \textbf{RQs} & \textbf{Tasks} \\
        \midrule
        \textbf{C1} & All tasks should reflect smart contract use cases that have received some attention in the blockchain community. & RQ1 & Auction, Prescription, Casino\\
        \textbf{C2} &  At least one open-ended task that could be used to evaluate whether Obsidian participants could create their own typestate-oriented interfaces (rather than only implementing according to a well-defined specification). & RQ3 & Casino \\
        \textbf{C3} & At least one task should put Solidity participants in a position where they might accidentally lose an asset. For Obsidian participants, the corresponding question is whether they are able to complete the task in spite of the strictures of ownership. & \parbox[t]{1cm}{RQ2,\\RQ3} & Auction, Casino \\
         \textbf{C4} & In a previous qualitative study of an early version of Obsidian, participants found it very challenging to use ownership (\citep{Coblenz2020:PLIERS}). Therefore, at least one task should assess whether the changes to Obsidian since its original design have addressed the difficulty of using ownership to restrict the use of assets. & RQ3 & Prescription, Casino \\
        \bottomrule
    \end{tabularx}
    \caption{Alignment between task design criteria and the tasks we designed.}
    \label{table:task-design-criteria}

\end{table}

The tasks and their results are described in detail in the following sections. Complete task materials are included in the supplement. Although we told participants that we might interrupt them eventually if they needed to move on to the next task, we reduced time pressure by not telling them specific per-task time limits.

\section{Auction Task}
\label{sec:auction}

 In the \textit{Auction} task, we asked participants to fill in missing code in an implementation of an English auction, in which bids are made openly and the highest bidder wins. To increase external validity (criterion C1 in \cref{table:task-design-criteria}), we modeled the task after an example from a Solidity tutorial \cite{solidity-auction}. We required that all bids be accompanied by funds to ensure that the winning bidder will pay for the item. When a bid is exceeded, the original bidder should receive a refund of their bid. We gave the participants 30 minutes to complete the task.
 
\Cref{obsidian-auction-task-code} shows the \code{bid} transaction that was provided to participants as well as a sample solution. In the first subtask, marked by \code{//~1.~TODO}, participants needed to write code to refund the existing bid to the previous bidder, whose address was stored in \code{maxBidder}, and record the new bid (\code{money}). In the second subtask (\code{//~2.~TODO}), participants needed to refund the bid to the bidder. The code in yellow shows a correct answer. In both cases, there was an opportunity for \textit{asset loss}: if participants overwrote the old \code{Money} reference (stored in \code{maxBid}), then the old bid would be lost. In Obsidian, the compiler would report an error if this happened; in Solidity, there was no protection against that mistake. This opportunity for error reflected task criterion C3.

We refine the high-level research questions for this task (using numbering that reflects the original research questions in \cref{sec:introduction}):

\begin{description}
	\item[RQ 2.1:] How frequently do Solidity participants accidentally lose assets in the Auction task?
	\item[RQ 3.1:] Overall, do completion times differ across conditions?
	\item[RQ 3.2:] Overall, are participants more likely to finish Auction (and do so correctly) if they use Obsidian rather than Solidity?
\end{description}

\subsection{Auction Results and Discussion}
% auction.overall <- matrix(c(2, 7, 7, 1), nrow=2, ncol=2, byrow=TRUE)
% fisher.test(auction.overall)
\Cref{auction-summary} summarizes the results of the Auction task; errors are shown in \cref{auction-errors}. Nine Solidity participants said they were done with the task before the 30 minutes expired; among these nine, the average time was 12 minutes (95\% CI: [6.8, 17.4]). Eight Obsidian participants said they were done with the task before running out of time; among these eight, the average time was also 12 minutes (95\% CI: [6.4, 18.3]). Thus, for \textbf{RQ 3.1}, the difference in times was not significant. Two participants completed the task correctly in the Solidity condition; seven completed the task correctly in the Obsidian condition. The difference in success rates (summarized in the first two rows of \cref{auction-summary}) is statistically significant, with $p \approx .015$ (Fisher's exact test; odds ratio 0.053). We conclude for \textbf{RQ 3.2} that participants who finished were more likely to finish correctly if they used Obsidian than if they used Solidity.

% solidity_error <- qt(.975, df=n_solidity-1)*sd_solidity/sqrt(n_solidity)
% obsidian_error <- qt(.975, df=length(obsidian_times)-1)*sd_obsidian/sqrt(length(obsidian_times))
\begin{table}[b]
\begin{tabular}{@{}lll@{}}
\toprule
& \textbf{Solidity} & \textbf{Obsidian} \\
\midrule
Completed task correctly	& 2 & 7 \\
Completed task with bugs & 7 & 1 \\
Time in min., completed tasks only; 95\% CI & 12; [6.8, 17.4] & 12; [6.4, 18.3]\\
Did not complete the task & 1 & 2 \\
\bottomrule
\end{tabular}
\caption{Auction task results. N=10 in each condition.}
\label{auction-summary}
\end{table}

\begin{table}[t]
\begin{tabular}{@{}lll@{}}
\toprule
& \textbf{Solidity} & \textbf{Obsidian} \\
\midrule
Ran out of time & 1 & 2 \\
Lost an asset in either subtask & 7 & 0 \\
\textit{Subtask 1} 	\\
\hspace{.5cm} omitted refund of old bid & 3 & 0 \\
\hspace{.5cm} overwrote old refund & 4 & 0 \\
\hspace{.5cm} refunded to wrong bidder & 0 & 1\\
\textit{Subtask 2} \\
\hspace{.5cm} overwrote old refund & 4 & 0 \\
\hspace{.5cm} refunded via \code{transfer()} instead of \code{pendingReturns} & 4 & N/A \\
\bottomrule
\end{tabular}
\caption{Errors in Auction task. N=10 in each condition.}
\label{auction-errors}
\end{table}

Of the two Obsidian participants who did \textit{not} finish the Auction task in time, one (P45) was confused about the semantics of the \code{::} field initialization operator, attempting to use \code{BidsMade::maxBid} to refer to the current value of the \code{maxBid} field rather than the future value after a state transition. This misconception led to a compiler error message that the participant did not find helpful. The other participant also received a confusing error message: although the code invoked a transaction that did not exist, the error message pertained to ownership of the transaction's parameter. A more mature compiler with better error messages might have helped the participants finish the task.

%P53
%The Solidity participant who did not finish the Auction task in time also did not finish either of the other two tasks in the study in time, and was the only participant who did not finish any of the three tasks. We suspect that this participant had weaker programming skills than the other participants, and that this was the primary reason for the outcome.

In \textbf{subtask 1} (starting at line 31 in \cref{obsidian-auction-task-code}), participants needed to record the new bid and refund the old bid. We found the following errors among the Solidity participants who said they were done:
\begin{enumerate}
\item Loss of previous refunds: the correct implementation \textit{added} the new refund to any prior refund. Four participants used \code{=} instead of \code{+=}, overwriting any old refund (in line 33).
\item Omission of refund: three participants neglected to refund the previous bid (e.g., omitting line 33).
\end{enumerate}

% auction.asset.loss <- matrix(c(7, 0, 2, 8), nrow=2, ncol=2, byrow=TRUE)
% fisher.test(auction.asset.loss)

All eight of the Obsidian participants who said they were done did so without losing any assets, since otherwise the compiler would have given an error. However, one participant refunded the old money to the \textit{new} bidder instead of to the \textit{previous} bidder. 

While doing the task, two of the Obsidian participants received a compiler error indicating that they had lost an asset. For example:

\begin{quote}
\code{auction.obs 37.28: Variable \textquotesingle maxBid\textquotesingle{} is an owning reference to an asset, so it cannot be overwritten.}
\end{quote}

Both of these participants successfully fixed the error.

In \textbf{subtask 2} (starting at line 41 of \cref{obsidian-auction-task-code}), participants needed to refund the new bid, since it was not larger than the previous bid. Among the nine Solidity participants who said they finished the task, two refunded the bid properly (using \code{pendingReturns}). Four refunded via \code{transfer}, which would not have resulted in asset loss but was inconsistent with the documentation we gave them. The documentation specified to use the withdrawal pattern to be consistent with the typical recommendation when using Solidity. Four attempted to refund via \code{pendingReturns} but, as in the first subtask, overwrote any previous refund, potentially losing money.

One might argue that the potential for asset loss due to improper use of \code{pendingReturns} was due to the need to use the \textit{withdrawal pattern}~\cite{Withdrawal-Pattern}, as discussed above. However, the particular bug we observed was due to participants overwriting an integer rather than adding to it, and we infer that arithmetic errors are likely common when manipulating assets manually. Obsidian protects against these bugs by encouraging programmers to design APIs that use assets to represent money rather than raw integers.

Of the seven Obsidian participants who completed the Auction task successfully, while they were working, two received compiler errors indicating that they had lost assets. One additional participant, P45, got an error about a lost asset, but did not finish the task because they were confused about the use of the \code{::} operator.

%find . -name *auction.obs.out | xargs grep "asset"
% P45, P47, and P50 all got errors about lost assets.
%P45 didn't finish (confused about ::). The resulting error is about ::.
%P47 refunded to the wrong bidder but didn't lose any assets.  But did fix the error.
%P50 finished successfully. Fixed compiler error to avoid losing the old bid.

We conclude (\textbf{RQ 2.1}) that asset loss was frequent among Solidity users, and \textit{more} frequent than among Obsidian users, who did not lose any assets ($p \approx .002$, Fisher's exact test). This difference may have been caused by a combination of differences in language design and differences in API design, but the different API designs were related to different language design choices.

\section{Prescription Task}
\label{sec:prescription}

To address task criterion C4, we gave participants a short \code{Pharmacy} contract (43 lines in Obsidian or 46 lines in Solidity including whitespace). The code included an example to show how the contract was vulnerable to attack. Although a \code{Prescription} was specified to only permit a fixed number of refills, a \code{Patient} could invoke \code{depositPrescription} on more than one \code{Pharmacy} object, resulting in the patient being able to refill the prescription the given number of times at \textit{each} pharmacy. We asked participants to fix the bug, avoiding runtime checks if possible. In Solidity, for example, \code{depositPrescription} had the signature below:
\begin{lstlisting}[language=Solidity]
function deposit(Prescription p) public returns (int);
\end{lstlisting}

In Obsidian, the starter code provided this signature:
\begin{lstlisting}
transaction deposit(Prescription@Shared p) returns int;
\end{lstlisting}

In Obsidian, it sufficed for participants to change the signature so that the \code{Pharmacy} acquired ownership of the prescription object:
\begin{lstlisting}
transaction deposit(Prescription@Owned >> Unowned p) returns int;
\end{lstlisting}

In Solidity, in contrast, since there is no static feature that would make the above safe, participants had to implement a global tracking mechanism across all \code{Pharmacy} objects. 

The task was based on a task from our prior study, which found that users of an earlier version of Obsidian had great difficulty using ownership to fix this problem~\cite{Coblenz2020:PLIERS}. For example, some participants in that study thought about ownership in a dynamic way (for example, writing \code{if} statements to test ownership) or were confused about when ownership was transferred between references. The version of the language used in the present study includes changes that resulted from that work, such as fusing typestate and ownership in the language syntax, making ownership transfer explicit in transaction signatures, and removing local variable ownership annotations. Our research questions were centered around evaluating the revised language:

\begin{description}
	\item[RQ 3.3:] What fraction of Obsidian participants could use ownership to fix the multiple-deposit vulnerability?
	\item[RQ 3.4:] Does using ownership to prevent the multiple-deposit vulnerability take less time than using a traditional dynamic approach?
%	\item[RQ 3.5:] Were participants using Obsidian more likely to finish (and to do so correctly) than participants using Solidity?
\end{description}

We gave participants 35 minutes to complete the task. Because ownership-based approaches are not checked by the Solidity compiler, we wanted to allow Solidity participants who proposed ownership approaches to try again. Therefore, when participants informed the experimenter that they were done, the experimenter inspected their code. If they had used static notions of ownership instead of a dynamic approach, participants were permitted to try again in the rest of their 35 minutes.

\subsection{Prescription Results}
%Obsidian
\Cref{table:prescription-summary} summarizes the results. Regarding \textbf{RQ 3.3}, six of the ten Obsidian participants successfully used ownership to solve the problem. The dynamic Obsidian solution that we judged to be correct tracked global state by making \code{Prescription} mutable, despite a comment indicating that \code{Prescription} should be immutable.

%Solidity
Five of the ten Solidity participants tried to use ownership, even though Solidity does not check ownership. Only three of the Solidity participants said that they were done within the time limit, and of those, only two had a correct solution. 
The incorrect Solidity solution attempted to solve the problem by making \code{Prescription} mutable to track remaining refills globally, but in addition, although the participant tried to track the number of refills across all pharmacies, the code did not update the global number of refills when refilling a prescription. 

\begin{table}[b]
\begin{tabular}{@{}lll@{}}
\toprule
& \textbf{Solidity} & \textbf{Obsidian} \\
\midrule
Attempted a static solution & 5 participants & 6 participants \\
Correct static solution & N/A & 6 \\
Attempted a dynamic solution & 6 & 3 \\
Correct dynamic solution & 2 & 1 \\
Made \code{Prescription} mutable & 2 & 1 \\
Completed within time limit & 3 & 9 \\
Mean time among successful participants; [95\% CI] & 20 min.; [0, 45] & 22 min. [12, 33] \\
\parbox[t]{7cm}{Mean time among successful participants after removing Solidity time spent on static attempts} & 18 min. & 22 min. \\
\bottomrule
\end{tabular}
\caption{Summary of Prescription task results. Times are shown as mean (standard deviation). N=10 in each condition. Two Solidity participants tried both static and dynamic approaches, and one Solidity participant made no changes, resulting in 11 Solidity attempts.}
\label{table:prescription-summary}
\end{table}

We separated time Solidity participants spent on static solutions from the time spent on dynamic solutions, since these attempts were likely prompted by the training materials; \cref{table:prescription-summary} shows both sets of times. However, we included all time spent by Obsidian participants on correct solutions because it is realistic that some Obsidian users would have tried each approach. 

Regarding \textbf{RQ 3.4}, we did not observe a significant difference in completion times. However, a large fraction of Solidity participants spent significant portions of their time attempting static solutions. In retrospect, it might have been more informative to have told Solidity participants explicitly that they needed to use a dynamic solution, and this approach might have led to a more interesting comparison. However, due to the small amount of code required for the static solution, we suspect that there is a significant learning effect to be leveraged here in Obsidian, and the participants who succeeded could likely do so again in a similar situation much faster. Furthermore, the fact that only two of three Solidity participants who finished this task did so correctly underscores the benefit of static enforcement of these kinds of safety properties. 

\subsection{Prescription Discussion}

One might have expected that applying ownership would be challenging, since the concept was new to the participants, but since only half of those who said they had completed a dynamic solution had correct solutions, it would appear that using the new static construct may not be harder than writing global state tracking code. In fact, using ownership to solve programming problems is \textit{teachable}: six of nine Obsidian participants who completed the task used ownership to do so. We expect that the remaining three could be taught to do so with additional practice.

In an earlier study~\cite{Coblenz2020:PLIERS}, which used a very similar task, four of six participants were able to solve the problem, but all of them received significant help from the experimenter. The remaining two did not solve the problem even with help. Although those numbers cannot be compared with the results of this study directly, the fact that six of our participants were able to complete the task without any help suggests that the changes since then have improved usability.

\begin{sloppypar}
Security experts have long argued in favor of immutable data structures~\cite{securecoding, seacord2013secure}, which is one reason why we specified that \code{Prescription} was immutable. However, these results point out that this approach may not be tenable: specifications of immutability may be ignored or removed, and attempts to maintain immutability require substantial work, which may itself be bug-prone. Indeed, when language-based mechanisms do not provide the required safety properties~\cite{Coblenz2017:Glacier}, it may be safer and cheaper to use a \textit{mutable} design than to bear the cost of immutability. In this case, making \code{Prescription} mutable would have obviated the need to implement separate data structures keeping track of how many refills each prescription had left, since that information could be kept in the \code{Prescription} object directly. That approach appeared to be more natural for the participants, and certainly required less code.
\end{sloppypar}

The benefits of Obsidian's ownership system in the Prescription task contrast with the benefits of other kinds of ownership. For example, ownership in Rust~\cite{rust}, which is understood to be one of the more challenging aspects of learning Rust~\cite{RustSurvey}, introduces constraints on mutation. However, in Rust, only owning references can mutate objects. In Obsidian, mutation is restricted only as much as is needed to provide sound typestate specifications, since concurrency is not a concern. Therefore, in Obsidian, only changing \textit{which named state an object is in} is restricted, and then only through \code{Unowned} references. This contrasts with Rust, in which \textit{all} modifications to fields of objects are restricted. Six of 10 participants were able to use the linear aspects of ownership alone in the Prescription task, suggesting that languages that adopt just linearity (and not mutability restrictions) may be usable. Perhaps by integrating a more flexible permissions system, languages such as Rust could be made more convenient for common cases, and thus have a more gradual learning curve. Alternatively, making these changes in Rust might come at the expense of expert efficacy, so the impact of these changes in the long term would need to be evaluated.

% Obsidian participants who didn't use ownership: P45, P56, P59, P61.
% P45 ran out of time on Auction and gave up on Casino
% P56 spent 2h28m on the tutorial, ran out of time on Auction, and didn't have enough time left for Casino (only 20m left)
% P59: no obvious excuse (talked about ownership but wasn't sure how to use it)
% P61: no obvious excuse. Ran out of time working on a dynamic solution.

Although six of the Obsidian participants used ownership successfully, four participants did not. One of the four participants did not complete any of the three tasks. Another ``fixed'' the issue by modifying code in \code{Patient} that we had provided as an example of how a nefarious patient might exploit the bug; perhaps this participant did not really understand the task. The other two seemed to need more time studying ownership in order to use it effectively. 

The instructions included: ``Please use what you have learned today to fix this problem (avoiding runtime checks if possible).'' Perhaps as a result, a similar fraction of participants tried to use ownership in both conditions. We wanted to encourage the Obsidian participants to use ownership so that we could assess to what extent they could use it effectively, but the instructions persuaded some of the Solidity participants to use it even though doing so was not checked by the compiler.

\section{Casino Task}
\label{sec:casino}

\begin{figure}[tb]
\includegraphics[width=.8\columnwidth]{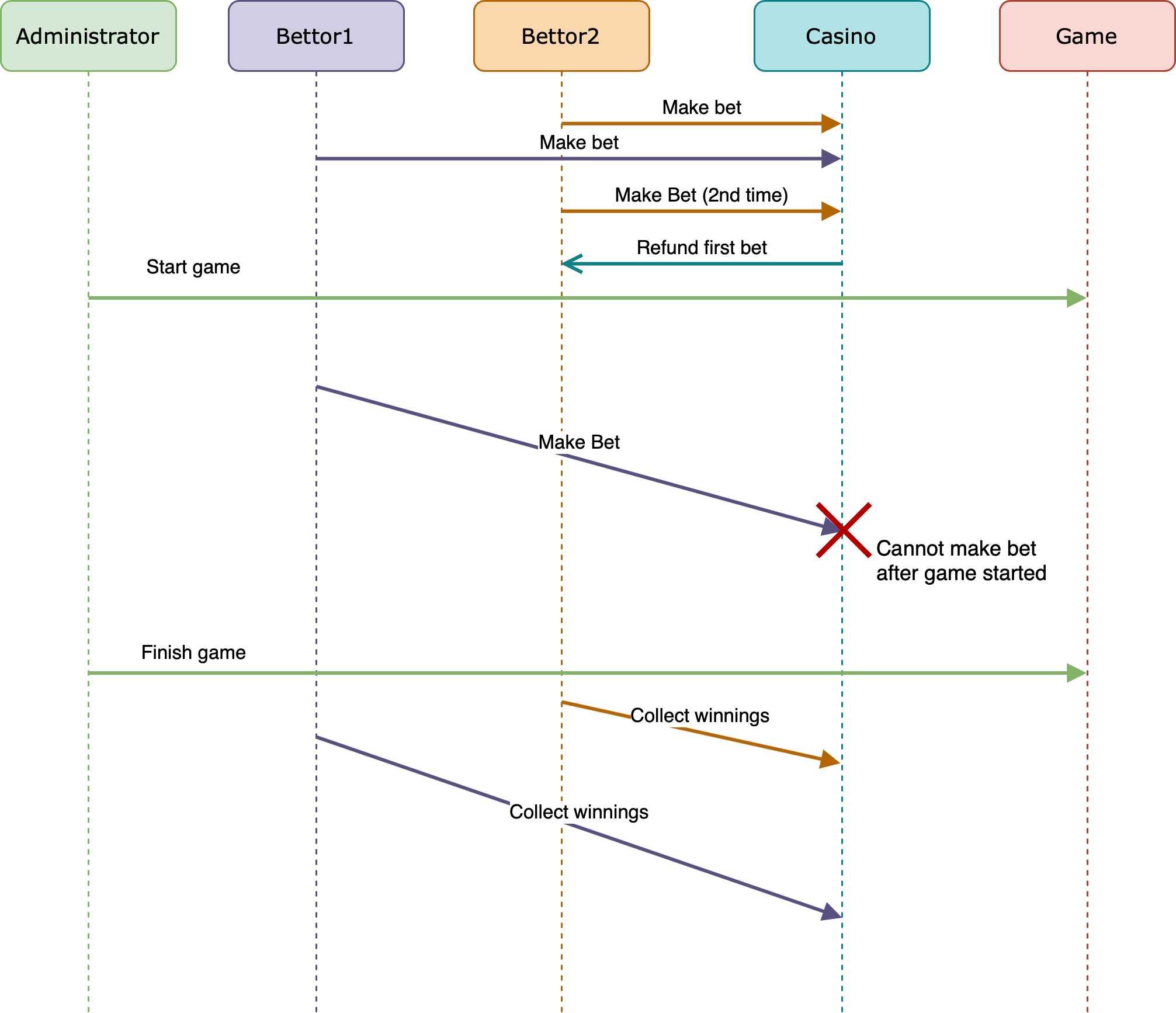}
\caption{Sequence diagram given to participants to show what operations the Casino contract should support.}
\label{casino-diagram}
\Description{The sequence diagram given to participants in the Casino task shows five entities: Administrator, Bettor1, Bettor2, Casino, and Game. Bettor1 and Bettor2 each make a bet to the Casino. Then, Bettor2 makes another bet, and the Casino refunds the first bet. Then, the Administrator starts the game by sending a message to the Casino. Bettor1 tries to make another bet, but the bet is disallowed because the game already started. The Administrator then finishes the game. Bettor1 and Bettor2 each collect their winnings.}
\end{figure}

The casino task was more open-ended than the other tasks, addressing criterion C2. We gave participants a web page with a diagram showing invocations that needed to be supported (\cref{casino-diagram}). The web page included a list of requirements:

\begin{enumerate}
\item If a \code{Bettor} predicts the outcome correctly, the \code{Bettor} gets twice the \code{Money} they put down. For example, if \code{Bettor} b puts down 5 tokens on the correct outcome, they should receive 10 tokens after the Game is played.
\item If the \code{Bettor} predicted incorrectly, the \code{Casino} keeps their tokens.
\item \label{bets-before-start} Bets can only be made before the \code{Game} starts.
\item \label{winnings-after-finished} Winnings can only be distributed after the \code{Game} is finished.
\item \code{Bettor}s must collect winnings themselves from the \code{Casino} after a Game by calling code, which you need to write. Until winnings are collected, the \code{Casino} keeps track of them.
\item A \code{Bettor} can have one active bet per game. If a \code{Bettor} bets more than once, their original bet should be replaced by the new one and any previous bet should be refunded.
\item A \code{Bettor} MUST put down tokens at the same time that they're making a Bet.
\item If the \code{Casino} does not have enough tokens available to pay out winnings, the invocation to collect winnings can fail.
\end{enumerate}

We provided starter code for \code{Casino}, \code{Game}, and \code{Bet}. Obsidian participants also received implementations of appropriate containers (Solidity has suitable built-in containers).

We used the Casino task to investigate five research questions:

\begin{description}
    \item[RQ 2.2:] How frequently do Solidity participants lose assets in the Casino task? 
	\item[RQ 3.5:] To what extent do Obsidian participants leverage typestate in transaction signatures to avoid dynamic checks (criterion C2)?
	\item[RQ 3.6:] In both versions, programs represented funds with \code{Token} objects. Does Obsidian's type system help participants avoid losing \code{Token} objects compared with Solidity (criterion C3)?
	\item[RQ 3.7:] Do Obsidian participants view \code{Token} objects as resources that should not be created or destroyed, or as data, which could be created and destroyed as needed (criterion C4)?
	\item[RQ 3.8:] How do task completion times compare between Solidity and Obsidian participants (criterion C4)?
\end{description}

%P58
Of the original ten participants in each condition, we excluded one Obsidian participant who should have received an error from the compiler but, due to a bug, did not.
%P47, P53, P56, P59, P64
We also excluded one Solidity participant and four Obsidian participants who did not have enough of their four hours remaining, and in the time available, were not satisfied with their solution. This left nine Solidity participants and five Obsidian participants whose results we analyzed. The results below pertain to these 14 participants. The discrepancy between the number of participants who had enough time across the two conditions was due to two factors: the faster task completion times by Solidity participants, and the high variance among Obsidian participants for time required for tasks that occurred \textit{before} Casino.

% Including for Obsidian: P45 (gave up), P50, P52, P61, P65
% P57 gave up in the Solidity condition.
Of the 14 participants whose results we analyzed, one participant in the Obsidian condition gave up after 1 hour, 15 minutes. That participant had chosen an unnecessarily difficult implementation strategy, requiring implementing a new container (implemented as a linked list). Also, the participant delayed trying to compile until after writing a lot of code, resulting in a large collection of compiler errors. The remaining four Obsidian participants all wrote code that compiled successfully. One participant in the Solidity condition gave up after 39 minutes, having received a parser error that they were not sure how to fix.

The Casino task results should be interpreted in the context of some additional limitations relative to the other tasks. The results of a comparison between conditions on this task may be biased because the Obsidian participants who were included for analysis in this task are those who completed the earlier tasks fastest. As a result, the Obsidian participants may have been stronger programmers on average than those in the Solidity condition (in which almost all the participants had time to try the task).

Casino was more open-ended than the other tasks, resulting in more variance, as participants made varying implementation choices. Given this, the small numbers, and the potential bias, we use this primarily as an opportunity to develop hypotheses, design insights, and identify future opportunities for improvement.

\subsection{Casino Results and Discussion}

We analyzed the code participants wrote, comparing the code to the requirements we gave them. As with the other tasks, in order to isolate the bugs from each other, our analysis was manual; the source code produced by each participant is included in the supplement. Results are summarized in \cref{table:casino-summary,table:casino-successes}. Before discussing the results for the four research questions, we discuss the errors that participants made while working on Casino.

\begin{table}[t]
\begin{tabular}{@{}lll@{}}
\toprule
& \textbf{Solidity} & \textbf{Obsidian} \\
\midrule
Had enough time to try Casino & 9 & 5 \\
Completed Casino with a program that compiled & 8 & 4 \\
\bottomrule
\end{tabular}
\caption{Summary of Casino task completions.}
\label{table:casino-summary}

\begin{tabular}{@{}lll@{}}
\toprule
& \textbf{Solidity ($N = 8$)} & \textbf{Obsidian ($N = 4$)} \\
\midrule
Completed task correctly (no identified bugs) & 12.5\% & 0\% \\
Winnings collection emits error if Casino is out of tokens & 62.5\% & 25\% \\
Casino keeps tokens when a bet is lost & 100\% & 50\% \\
%Bets only permitted before game starts & 87.5\% & 100\% \\
%Winnings only distributed after game finishes & 87.5\% & 100\% \\
Bettor's extra bets result in refunds & 100\% & 75\% \\
Only used disown safely & N/A & 0\% \\
Managed tokens correctly (not fabricating or losing them) & 50\% & 0\% \\
Mean completion time & 37 min. & 64 min.\\
\bottomrule
\end{tabular}
\caption{Summary of Casino task results among completed programs that compiled, showing correct solution rates among errors made by more than one participant.}
\label{table:casino-successes}
\end{table}

Except for the one Solidity participant who completed the task correctly, all of the other participants across both conditions inserted various bugs.  For example, three participants in each condition failed to emit an error when attempting to collect winnings from the Casino when the Casino lacked enough resources to distribute them.

The most common bug among Solidity participants was some kind of token loss or improper fabrication; four of the eight made this mistake, addressing \textbf{RQ 2.2}. For example, one participant, when accepting a new bet, first credited the bettor's account for any prior bet, and then returned if the bet amount exceeded the bettor's balance. This meant that a second bet, when disallowed, would incorrectly fabricate tokens out of thin air, since the code that debited the bettor's balance occurred after the early \code{return}. Another participant neglected to debit accounts for extra wagers, also fabricating tokens out of thin air. This shows that protection against incorrect asset manipulation \textit{is} important.

%%%%%%%%%% Usage of disown %%%%%%%%%%
% P50, P52, P61 and P65
All four Obsidian participants who finished the Casino task used \code{disown} improperly to throw away assets. We found this surprising, since we had warned the participants against improper use of \code{disown}. The tutorial included an example of how \code{disown} might be needed inside the implementation of a \code{Money} contract, and wrote below the example:
\begin{quote}
IMPORTANT: \code{disown} should be used only when you really want to throw something out. Above, \code{disown} is required because of the manual arithmetic used to manipulate \code{amount} inside the implementation of \code{Money}, but it is not needed in most normal code.
\end{quote}

It would not have sufficed to remove \code{disown} from the language; in addition to the fact that it is needed in certain (rare) cases, programmers could build a \code{Trash} contract to hold discarded objects, thus suppressing any errors the compiler would emit. We have several hypotheses regarding why \code{disown} was abused:

\begin{itemize}
	\item Some participants may have used \code{disown} to silence the compiler when they felt they had a correct solution. One participant discarded the old wager, using money from the casino's pot to pay out bets. The participant also disowned the bet when a losing bettor tried to retrieve their winnings (a correct solution would have put the tokens in the casino's pot). %P50
	\item Some participants may not have read or understood the tutorial's warning about \code{disown}; in retrospect, we should have assessed understanding explicitly.
	\item Some participants did not sufficiently understand the notion of assets. For example, when disbursing winnings, one solution disowned the previous wager and created new \code{Token}s when needed, rather than reusing the tokens from the wager. %P61
	\item Some participants may have used \code{disown} as a workaround for an unsolved problem. In code to accept a new bet, one participant disowned any previous bet by that bettor, and then wrote: \code{// Currently just throws the bettor's money away and hopes they find it eventually}. %P65
\end{itemize}

This motivates a question for future research to characterize the use of these \textit{escape hatch} constructs, which can be used in both safe and dangerous ways. Some languages include warnings in the names of such constructs, as in \code{unsafePerformIO} in Haskell~\cite{UnsafePerformIO}, but such approaches may not be effective in explaining the danger. 

\textit{Risk compensation} refers to the idea that people compensate for safety features by taking additional risks. For example, drivers may drive faster when wearing seat belts because the seat belts may mitigate the risk caused by higher speeds. However, in many cases (such as in the seat belt example), overall risk is reduced despite potential compensating behavior~\cite{Houston2007:Risk}. Further study is needed to consider the question of whether risk compensation occurs with strong type systems. Perhaps some participants who used \code{disown} assumed that if there were a bug, the compiler would report it.

%%%%%%%%%% Usage of typestate %%%%%%%%%
Regarding \textbf{RQ 3.5}, the \code{Game} contract that we provided defined states for \code{Game} (\code{BeforePlay}, \code{Playing}, \code{FinishedPlaying}). When implementing \code{Casino}, participants wrote code in their \code{Casino} transactions (to implement requirements \ref{bets-before-start} and \ref{winnings-after-finished}) to dynamically check the state of the \code{Game}. An alternative approach, which we had expected some participants to use, would have involved observing that the state of \code{Game} was related to the state of \code{Casino}. If a \code{Casino} was in a state that permitted bets, then the \code{Game} must not have started. Likewise, collecting winnings was only possible from a \code{Casino} whose \code{Game} had ended. Our starter code defined states in \code{Game}. Therefore, participants could have avoided dynamic checks in \code{Casino} by defining multiple states in \code{Casino}, each of which corresponded to a particular state of \code{Game}. All of the participants added the dynamic checks. The participant who gave up tried to check the state with a static assertion, evidently not understanding that the assertion was static, not dynamic. 

The lack of usage of static state information is unfortunate because it represents a missed opportunity to rule out bugs in calling code. Perhaps more-experienced programmers would be more interested in leveraging this language feature; alternatively, it might require more training or a more-convenient language design. The design the participants chose may have been best given their incentives; the typestate-based approach would have required adding more structure to reflect the typestate relationships between \code{Casino} and \code{Game}. However, motivated by the results of this study, we hope to consider future language design changes that make typestate coupling of different objects convenient.

Perhaps \textit{creating} new interfaces that use novel verification-related features (such as typestate) is harder than \textit{consuming} them, in which case further research should consider novel ways of scaffolding interface design and creation. In Obsidian, it is possible to use states to define different fields (each with its own typestate specification) for each possible state of referenced objects. This approach would couple the states of the two objects, letting the programmer avoid runtime checks that would be otherwise required, but we did not train participants in this approach. Perhaps this technique is sufficiently subtle that we should have provided training directly.

Surprisingly, in \textbf{RQ 3.6}, we observe that in fact, Solidity participants were probably more likely to correctly have the casino keep tokens when a bet is lost (\cref{table:casino-successes}, $p \approx 0.09$, Fisher's exact test). Likewise, Solidity participants may be more likely to successfully issue refunds for bets after the first bet ($p \approx 0.33$, Fisher's exact test). We believe this is related to abuse of \code{disown} by Obsidian participants.

% Used tokens as assets: P45
% Used tokens as data (either created or destroyed them): P50, P52, P61 (new Tokens), P65.

Regarding \textbf{RQ 3.7}, all four of the Obsidian participants who wrote solutions that compiled treated tokens as data rather than assets, i.e., at some point in their code, they either created new tokens or disowned tokens. The Obsidian participant who gave up did not do either of those things, likely viewing tokens more as assets, but became mired in a list of type errors. We conclude that use of assets was likely \textit{not} natural for our participants. This interpretation is consistent with the post-study survey results (\cref{sec:post-study-survey}), in which we observed that participants said they felt ownership and states were more useful than assets.

For \textbf{RQ 3.8}, Obsidian participants spent significantly longer on the Casino task than Solidity participants did ($p \approx 0.02$, Mann-Whitney U test, $d \approx 1.9$). Therefore, the stronger type system provided by Obsidian likely had a significant cost in development time for our participants. We hypothesize that this cost is greater with more open-ended tasks, which would explain why we did not observe this difference in the two prior tasks. Of course, the additional cost may be worthwhile since Obsidian would rule out some classes of bugs statically, particularly when used by skilled programmers who do not abuse \code{disown}. The time to task completion in this experiment might relate in practice to the time required to complete a \textit{prototype} version of software rather than the time required to create a production-quality version.

%% P50, P52, P61, and P65 are the only ones who finished Casino successfully.
% P50: makeBet(Bettor@Unowned bettor, Bet@Full >> Unowned bet)
% P52: makeBet(Casino@Shared this, Bettor@Unowned bettor, Bet@Full >> Unowned bet)
% P61: placeBet(Tokens@Owned >> Unowned bet, int prediction, Bettor@Owned bettor)
% P65: placeBet(int bettor_id, Bet@Full >> Unowned wager)

% Did anyone make states of Casino that matched Game?

% P50, P52, P61, P65: no
% P45, who gave up: no.

%% Why did Obsidian users fail to let Casino keep wagers when bettors lose?
% P52: disowned wager.
% P61: disowned wager unconditionally.

\section{Post-study Survey Results}
\label{sec:post-study-survey}

We conducted a post-study survey asking participants about their opinions. We asked participants (on a 5-point scale) how well they understood particular concepts and how useful they thought those concepts were. The results from participants who completed the survey \textit{before} being de-briefed from the study are summarized in \cref{survey-summary}. Although the Obsidian participants said they thought ownership was significantly more useful than the Solidity participants did ($p \approx 0.002$, $d \approx 2.5$), the Solidity participants indicated that they felt they understood states better than the Obsidian participants did ($p \approx 0.04$, $d \approx 1.3$). Perhaps the existence of an unfamiliar \code{state} construct in Obsidian, or the unfamiliarity of the relationship between states and types, led to less confidence. There was no significant difference in views of the utility of states. This may be due to the tasks we gave, which did not particularly rely on the static aspects of states.

\begin{table}[t]
\begin{tabular}{@{}lll@{}}
\toprule
& \parbox[t]{1.1cm}{\textbf{Solidity\\(N=6)}} & \parbox[t]{1.3cm}{\textbf{Obsidian\\(N=8)}} \\
\midrule
How much did you like the language you used? & 3.7 (0.82) & 4.0 (0.53) \\
How well do you feel you understand the concept of ownership? & 3.8 (0.98) & 3.75 (0.99) \\ 
*How useful do you think ownership is? & 3.0 (1.1) & 4.88 (0.36) \\ % p = 0.0017
*How well do you feel you understand the concept of states? & 4.8 (0.41) & 4.1 (0.64) \\ % p = 0.043
How useful do you think states are? & 4.3 (0.81) & 4.1 (0.64) \\
How well do you feel you understand the concept of assets? & 3.2 (0.98) & 3.4 (1.3) \\
How useful do you think assets are? & 2.7 (0.52) & 3 (1.2) \\
\bottomrule
\end{tabular}
\caption{Perceptions of ownership, states, and assets on a 1--5 scale (5 is best). Cells show average (standard deviation). *~indicates that a Mann-Whitney U test shows a significant difference at $p < 0.05$.}
\label{survey-summary}
\end{table}

We also compared across questions. Obsidian participants said that they felt both ownership and states were \textit{more useful} than assets ($p \approx .0027$, $d \approx 2.1$ and $p \approx .037$, $d \approx 1.2$, respectively, according to a Mann-Whitney U test and Cohen's $d$). The differences between perceptions of understanding between ownership and assets and between states and assets were not significant at $p < .05$. Of course, perceptions of understand and utility were influenced by the particular tasks the participants did and their perceived success at doing those tasks, but these results correspond with the Obsidian participants' failure to regard tokens as assets in the Casino task.

The survey also included a free-form box in which participants could respond to the question ``Please use this space to write any additional comments you have about the language or the study.'' We reviewed the free-form comments, and describe the most interesting ones here. Due to the sparse and brief nature of the data, we did not conduct a more formal analysis. The complete data are included in the supplement. 

Several of the Solidity participants expected that the language would have included constructs for states or ownership. For example, one participant wrote in the post-study survey:
\begin{quote}
It also seemed like there should be some syntactic sugar for writing things like:
\begin{lstlisting}
enum State { Foo, Bar, Buzz }
State s
\end{lstlisting}
since they are so common.
\end{quote}

Three Solidity participants expected that the compiler would check ownership. For example:

\begin{quote}
On [semantics] --- I was hoping ownership / assets / states would be statically verified. When I wrote code during the 2nd phase of the study, I found that I didn't really document ownerships / asset status. 
\end{quote}

Similarly, from another Solidity participant:
\begin{quote}
I think it would be nice to have some static analysis to check ownership information rather than relying on the programmer to have good comments documenting ownership because in practice documentation is never perfect and often overlooked.
\end{quote}

Participants using Obsidian had differing opinions regarding how easy it was to learn. One wrote: 
\begin{quote}
The tutorial and the exercises are well-written and they helped me a lot in understanding the concepts of new language!
\end{quote}

Another Obsidian participant wrote:
\begin{quote}
The smaller coding exercises were nice to follow and complete. The open-ended part was a little overwhelming to finish, for somebody that just got introduced to new concepts of ownership, states and assets.
\end{quote}

One Obsidian participant commented on how ownership seemed natural after some practice:
\begin{quote}
\ldots the general concepts of ownership [were] a little unintuitive but after working with the language they started to make more sense and seem more natural\ldots.
\end{quote}

\section{Summary of Results}
\label{sec:results-summary}

In order to summarize the overall results across the three tasks, we computed the fraction of successfully completed tasks for each participant, taking into consideration the tasks that were not included in the analysis above (only nine Solidity participants and five Obsidian participants in the Casino task). The median fraction of tasks completed by Solidity participants was 0\%; the median fraction of tasks completed by Obsidian participants was 67\% (since these are ratios, we do not take their mean). We compared these fractions using a Mann-Whitney U test, observing a significant difference with $p < 0.008$. \Cref{table:task-completions} summarizes \textit{successful} task completions: tasks that were completed correctly within time limits.

%W = 84. Conducted test in R:
% obsidian_task_completions <- c(0.3333333333, 0.5, 0.6666666667, 0.6666666667, 0, 1, 1, 0.3333333333, 1, 0.6666666667)
% solidity_task_completions <- c(0, 0, 0, 0.3333333333, 0, 0, 0, 0, 0.3333333333, 1)
% wilcox.test(obsidian_task_completions, solidity_task_completions)

\begin{table}[tb]
    \centering
    \begin{tabular}{p{3cm} p{2cm} p{2cm}}
    \toprule
        \parbox[t]{3cm}{\textbf{Tasks completed\\correctly}} & \parbox[t]{2cm}{\textbf{Solidity\\ participants}} & \parbox[t]{2cm}{\textbf{Obsidian\\ participants}} \\
    \midrule
        0 & 7 & 1 \\
        1 & 2 & 3 \\
        2 & 0 & 6 \\
        3 & 1 & 0 \\
    \bottomrule
    \end{tabular}
    \caption{Numbers of participants completing different numbers of tasks correctly in the two conditions.}
    \label{table:task-completions}
\end{table}

We now return to the initial research questions. 

\begin{quote}
    RQ1: Could we obtain actionable data about the usability of a novel programming language (that uses a type system that would be unfamiliar to our participants) in a short-duration user study (less than one day), which would be representative of real-world smart contract development?
\end{quote}

Our results regarding the other research questions show that we were able to identify both strengths and weaknesses of Obsidian relative to Solidity in the context of smart contract programming tasks that were drawn from real-world scenarios. Perhaps more importantly, the study led to insights about the type system and its potential risks that can be used to refine future language designs. For example, strong type systems can be learned and used to significant benefit in short periods of time, but escape hatches can be frequently abused.

\begin{quote}
    RQ2: Do programmers using Solidity insert more of the kinds of bugs that Obsidian is designed to catch?
\end{quote}

In Auction, we observed that Solidity participants lost assets frequently (seven of nine Solidity participants who finished lost assets) (RQ 2.1). Similarly, four of eight Solidity participants lost assets in the Casino task (RQ 2.2). We conclude that linear type systems likely have value in helping programmers of smart contracts detect bugs earlier. However, the abuse of \code{disown} in Casino shows that training or language refinement may be needed to help users of linear type systems obtain these guarantees effectively.

\begin{quote}
   RQ3: Can programmers who were previously familiar with object-oriented programming (but not with Obsidian, typestate, ownership, or linear type systems in general) successfully use Obsidian to complete relevant smart contract programming tasks? If so, is there a significant impact on task completion times?
\end{quote}

In Auction, we observed no significant difference in completion times, which is promising for strong type systems (RQ 3.1). Furthermore, participants who used Obsidian were more likely to finish Auction correctly (RQ 3.2). In Prescription, 60\% of Obsidian users were able to use ownership to fix the vulnerability we gave them, suggesting that earlier work iterating on the language design using user-centered methods may have been effective in making the language more usable than it had been initially (RQ 3.3). Using ownership does not necessarily take less time than using a dynamic approach, but seven of the ten Obsidian participants successfully completed the Prescription task, suggesting that enabling use of ownership can empower programmers to be more effective (compared to two of ten Solidity participants who succeeded) (RQ 3.4). In Casino, we observed that all of the Obsidian participants who finished the task abused the \code{disown} keyword, which serves as a caution that escape hatches from safety features are easily misused; this misuse can significantly hamper a language's ability to achieve its safety goals (RQ 3.5). Likely due to this misuse, Solidity participants were more likely than Obsidian participants to have the Casino keep funds from losing bets and to issue refunds correctly for revised bets. Likewise, the Obsidian participants appeared to create and destroy tokens at will, rather than treating them as assets; more training is likely required if we want programmers to achieve the safety benefits of adopting this perspective, which may not be a particularly natural one for them.  Finally, Obsidian participants spent significantly longer on the task than the Solidity participants did, confirming that in some more complicated tasks, a stronger type system likely increases the cost of an initial implementation.

% \begin{table}[ht]
%     \centering
%     \begin{tabular}{p{6cm}p{6cm}l}
%     \toprule
%     \textbf{High-level research questions} & \textbf{Task-based research questions} & \textbf{Conclusions} \\
%     \midrule
%      RQ1: Could we obtain useful data about the usability of a novel programming language in a short-duration user study? \\
%      RQ2: Do programmers who use Solidity accidentally insert the kinds of bugs that Obsidian is designed to detect? \\
%      	& RQ 2.1 How frequently do Solidity participants accidentally lose assets in the Auction task? \\
%      RQ3: Can participants successfully use Obsidian to complete relevant smart contract programming tasks? If so, is there a significant impact on task completion times? \\
%         & RQ 3.1: Overall, do completion times differ across conditions? \\
% 	    & RQ 3.2: Overall, are participants more likely to finish Auction (and do so correctly) if they use Obsidian rather than Solidity?\\
% 	    & RQ 3.3: What fraction of Obsidian participants could use ownership to fix the multiple-deposit vulnerability? \\
% 	    & RQ 3.4: Does using ownership to prevent the multiple-deposit vulnerability take less time than using a traditional dynamic approach? \\
%      \bottomrule
%     \end{tabular}
%     \caption{Summary of all research questions.}
%     \label{tab:result-summary}
% \end{table}

There has been long-standing debate about a hypothesis that dynamically-typed languages may be better for prototyping work than statically-typed languages~\cite{Hanenberg2014,Meijer2004:Static}. Our results provide a limited form of support for this hypothesis, since the Solidity participants were able to finish the Casino task faster than the Obsidian participants. Of course, since there were many differences other than type system between the two languages, the study does not address this hypothesis directly.

\section{Limitations and threats to validity}
The pre-screening instrument may have introduced bias, either by being unrepresentative of common content knowledge among typical object-oriented programmers or by discouraging some people from participating. The student participants may not be representative of the population of smart contract programmers. However, since most of the students had some professional experience, they were likely representative of entry-level programmers in industry \cite{StackOverflow2019:Survey}. Also, other studies have found no significant differences in code correctness in programming studies between students and non-students~\citep{Acar2017:Security}. The tasks were more constrained than real-world programming tasks, although smart contracts tend to be small in practice, averaging 322 lines \cite{Pinna2019:Massive}. \Cref{program-lengths} describes solution lengths in our study.

\begin{table}[b]
\begin{tabular}{@{}lllll@{}}
\toprule
& \multicolumn{2}{c}{\textbf{Solidity}} & \multicolumn{2}{c}{\textbf{Obsidian}} \\
& Min & Max & Min & Max \\\midrule
Auction & 291 & 355 & 352 & 395 \\
Prescription & 455 & 570 & 457 & 518 \\
Casino & 257 & 425 & 135 & 416 \\
\bottomrule
\end{tabular}
\caption{Ranges of all (whether correct or not) solution lengths in lines of code.}
\label{program-lengths}
\end{table}

The participants were new to the programming languages and to smart contract development in general, so it is possible that experienced programmers would have behaved differently. Although we tried to infer how particular aspects of the languages and their type systems affected participants' behavior and performance, because this was a summative study, the results may have been influenced other aspects of the experience, such as the way in which different aspects of the type systems interacted or the details of the particular tasks we gave participants.

The order of the tasks was consistent among all the participants, so the results of the tasks cannot be regarded as being independent of each other due to learning effects. Because of the way we allocated time, only five Obsidian participants completed the Casino task, so those results should be considered exploratory. 

We assessed results by manually analyzing code for correctness rather than by test cases. It is possible that using unit tests would have revealed additional bugs that we did not identify. Furthermore, disallowing execution might have biased the results to favor Obsidian, since without testing, writing correct code may be easier with the stronger type system that Obsidian provides.

Providing unit tests to participants (and allowing them to be run) might have allowed the participants to identify minor bugs that had gone undetected without incurring the time and variance cost of having participants write their own tests. This approach would also likely have increased external validity, although it adds a concern that participants might simply iterate until the tests pass rather than ensuring that they really understand how to write the code correctly. Furthermore, given our research focus on \textit{static} correctness techniques, it is not clear which unit tests should have been included.

Our study was potentially subject to the Hawthorne effect, in which participants may change their behavior when they know they are being observed~\cite{Sadler1996:Evaluating}. However, this effect would have been equally present in both conditions. To minimize this effect, although the experimenter remained in the room, the experimenter minimized direct observation by recording screen videos and source code rather than continuously watching participants. 

\section{Implications on PLIERS}

PLIERS is a process for language design that integrates user-centered methods into the design and evaluation process for programming languages~\cite{Coblenz2020:PLIERS}. In addition to evaluating Obsidian, the study also served in part to evaluate the PLIERS process. We were able to show a safety benefit of Obsidian in the Auction task, and were able to show that most of the Obsidian participants were able to use ownership successfully in the Prescription task. This shows that the tutorial method was mostly successful (though more success could likely have been obtained with more practice) and that the language design was effective overall (modulo the abuse of \code{disown} that we observed). Every study design involves making tradeoffs. The results here may show a tradeoff between training time and success rates; users of PLIERS will need to decide, based on their own design and research goals, how to balance the risks when designing their studies. However, the overall PLIERS design process did result in a language that had significant benefits relative to the status quo, which we were able to measure in a relatively low-cost study.

In retrospect, since only one of 20 participants completed the Casino task successfully (across both conditions), that task was too hard for the amount of time we allowed. We recommend that users of PLIERS carefully select success criteria in pilot studies in order to set appropriate task time limits and difficulties.

Designing and executing an effective RCT is extremely challenging and labor-intensive. The tutorial and tasks were initially drafted much earlier than the RCT, but even with a six-participant usability study (which had nine pilots) and four pilots for the RCT, we were still surprised at how difficult the Casino task was for some participants. In addition, executing the RCT was a lengthy effort; we estimate that recruiting participants, managing 21 participants plus four pilots, analyzing the data, etc. took about two months.

\section{Related Work}

Rust~\cite{rust-ownership} supports a version of ownership. Permissions and typestate features have not been included in popular programming languages, so we lack data regarding their broader usability. We are not aware of prior quantitative user studies of any of these kinds of type systems.

Other programming languages were designed to improve smart contract safety. Flint~\cite{Schrans2019:Flint} is a typestate-oriented language that supports linear assets, but omits a permissions system for flexible referencing of objects. Pact~\cite{Pact} is Turing-incomplete, avoiding nonterminating behavior. Scilla~\cite{Sergey2019:Safer} is an intermediate language whose semantics were formalized in Coq; it represents programs as communicating automata, avoiding complex inter-contract transactions (instead requiring that these be implemented as continuations). Nomos~\cite{Das2019:ResourceAware} uses linear types for safety and provides automatic resource analysis to facilitate gas usage prediction, but does not operate on any blockchain platforms. None of the above languages were evaluated empirically.

\citet{Delmolino2016:Step} described a user study of Serpent, which was a precursor to Solidity, showing several classes of bugs that occurred in the lab. Among these was \textit{asset loss}, which motivated our study to see whether our participants would avoid these bugs when using Obsidian.

In the past, we argued for using a variety of methods when designing programming languages~\citep{Coblenz18:Interdisciplinary}. \citet{Stefik14:Programming} focused on the need for empirical evaluation of programming languages. Stefik et al. developed methodology to evaluate syntax choices for novice programmers~\cite{Stefik2013:Empirical}. Hanenberg et al. compared static and dynamic type systems~\cite{Hanenberg2014}, finding benefits of static type systems in both documentation and compile-time checking. The general question of the benefits of the Obsidian type system relates to the more general question of static vs. dynamic types, which has also been addressed in other contexts~\cite{Hanenberg2014}. Our evaluation approach is related to that used by Stefik, Hanenberg, and others, but our method integrates teaching a novel type system into the experimental procedure. Also, our study was between-subjects, whereas the experiment comparing static to dynamic types was within-subjects~\cite{Hanenberg2014}. Our design is robust to learning effects, but more subject to variance among participants. In comparison to the Hanenberg et al. paper, we focus more on an experimental design that minimizes time required by participants and on reporting detailed data on errors made by participants.

Uesbeck et al. evaluated the benefit of C++ lambdas~\cite{Uesbeck:2016:ESI:2884781.2884849}, finding that no benefit even for the purposes for which lambdas were created. The only other quantitative empirical studies we are aware of for complete, novel programming languages (as opposed to ones that were already familiar to the participants) were of Quorum~\cite{Stefik2013:Empirical} and HANDS~\cite{Pane2002:Using}. Other work has focused on language extensions, such as for software transactional memory~\cite{Pankratius2014:Software} and immutability~\cite{Coblenz2017:Glacier}.

\begin{sloppypar}
Several languages have integrated support for linearity. Wadler~\cite{Wadler90:Linear} proposed the use of linear types for programming languages. In \textit{functional} languages, linearity may take the form of session types~\cite{Caires10:Session}. This approach mirrors typestate, since channel types (expressed as session types) change as messages are sent through them. Typestate was originally proposed by Strom and Yemini~\cite{strom1986typestate}. \citet{Drossopoulou2002:Fickle} introduced typestate for object-oriented languages in Fickle. In Fickle, objects could dynamically change class, but the static type did not reflect the changing classes. More recent work by DeLine, Aldrich, Bierhoff, and others describes how object-oriented languages can be used with typestate-specifying references~\cite{DeLine2004, Aldrich09:Typestate, Bierhoff:2007:MTC:1297027.1297050}. \citet{Garcia:2014:FTP:2684821.2629609} gave a formalization of typestate. None of the these languages were empirically evaluated with users, although \citet{Sunshine2014} showed that typestate in documentation can be beneficial. We designed Obsidian~\cite{Coblenz2020:Obsidian} using user-centered design~\cite{Coblenz2020:PLIERS}; that paper focuses on the design methodology, rather than on an empirical comparison between Obsidian and Solidity, but it describes a partial, previous version of the experimental materials that we used here.
\end{sloppypar}

High-quality error messages are a key component of a usable compiler. Some work in this area has produced guidelines for creating error messages that are effective for novices~\cite{Becker2019:Compiler}. Our work focuses on professionals, but the guidelines given in that work (for example, \textit{show solutions or hints}) are relevant to Obsidian.

\section{Future Work}
Casino, which included significant use of nominal states (i.e., statically-defined states with their own names), resulted in Obsidian participants writing dynamic checks. Future work should investigate the extent to which typestate, as provided in Obsidian and other typestate-oriented languages, can be made more compelling for programmers. For example, when pairs of objects have states that are coupled, the language could provide features to make representing and using this relationship convenient. Likewise, the existence of risk compensation among programmers should be investigated in future studies.

This study investigated whether Solidity programmers insert bugs that Obsidian detects, but a corpus study could show how prevalent these kinds of bugs are in real-world Solidity code. Bugs involving asset loss can occur in any language in which programs may manipulate assets; a corpus study involving a larger corpus of programs might be fruitful and still generalize to smart contract development. Indeed, it may be that the kinds of safety properties needed in smart contract development are generally applicable to many kinds of programs, regardless of whether they are hosted on blockchains.

A study that included testing, debugging, and code review would be more representative of real-world use. Also, the participants were new to the language they used in the study, and the Casino task may have been too difficult for the amount of time we allocated. A longer-duration study would likely have increased validity --- both by allowing enough time for difficult tasks and by allowing participants to become more comfortable with the language they were using. In a study of experienced Obsidian and Solidity programmers, we hypothesize that the task completion time difference would diminish significantly but Obsidian users would continue to take longer on complex development projects. On the other hand, there would likely be fewer serious bugs in the completed Obsidian projects, since Obsidian rules out classes of important bugs.

The tradeoff between internal and external validity seems fundamental to programming studies. Above, we suggested increasing external validity, but doing so might compromise internal validity further, since the tasks would include more kinds of work. Another approach to study design might trade external validity for additional internal validity. Asking participants to write nontrivial programs, such as in the Casino task, may reflect real-world programming tasks, but the task complexity results in high variance and significant amounts of time spent on programming issues that are not necessarily directly related to the research questions. In future studies, it might be worthwhile to consider more restricted tasks that only investigate a small portion of the development workflow. For example, if the research question pertains to creation of interfaces or architectures, then the task might isolate that part of the programming process rather than integrating it into a complete programming problem. Another approach to increasing external validity would be to investigate whether the results generalize to other languages or contexts. Do Java programmers who write auction programs also tend to accidentally lose assets? How does task performance compare between Obsidian and other linearly-typed languages, such as Nomos~\citep{Das2019:ResourceAware}?

Some of the usability results for Obsidian may generalize to other languages that use ownership, such as Rust. Future work should investigate whether the usability tradeoffs we observed here occur in other languages.

We observed many Obsidian programmers making unsafe use of \code{disown}. Future research should investigate how to more safely provide features that are safe in unusual situations but \textit{unsafe} in situations that arise commonly. Some languages use naming conventions, such as ``unsafe,'' to denote features that defeat the language's type system (e.g., Haskell's \code{System.IO.Unsafe.unsafePerformIO}), but here, the feature is a required part of the type system rather than a way of escaping from it.

\section{Conclusion}
As programming languages are tools for empowering programmers, empirical methods offer an opportunity for designers to provide evidence of the benefits of their work. In this study, we showed that programmers who used Obsidian were able to complete more of the tasks correctly than those using Solidity (\cref{sec:results-summary}). We also showed that ownership alone can be used effectively with a short training period and that assets can be used to detect bugs that would otherwise likely be inserted. However, we also identified areas of risk relating to language features that weaken the checks that the compiler provides. Although few language designs have been evaluated in this way, our work shows that it is possible to empirically evaluate a novel language, to both support hypotheses of usability as well as identify areas for potential improvement. We also hope that our findings of usability for the less-common type system features we analyzed will lead to more adoption of safer, more sophisticated type systems in future languages.

\begin{acks}
This research was sponsored by the \grantsponsor{NSA}{National Security Agency Department of Defense}{https://www.nsa.gov} award \grantnum{NSA}{H9823018D0008}; by the \grantsponsor{NSF}{National Science Foundation}{http://nsf.gov/} awards \grantnum{NSF}{CNS1423054} and \grantnum{NSF}{CCF1901033}; by the \grantsponsor{USAF}{United States Air Force Office of Scientific Research}{http://www.wpafb.af.mil/afrl/afosr/} award \grantnum{USAF}{FA8702-15-D-0002}; by two \grantsponsor{IBM}{IBM}{https://www.ibm.com/} PhD Fellowship awards; and by \grantsponsor{Ripple}{Ripple}{https://www.ripple.com}. The views and conclusions contained in this document are those of the author and should not be interpreted as representing the official policies, either expressed or implied, of any sponsoring institution, the U.S. government or any other entity.
\end{acks}

\bibliographystyle{ACM-Reference-Format}
\bibliography{obsidian.bib}

\appendix

\end{document}